\documentclass[twocolumn]{aastex631}
\usepackage[english]{babel} 
\usepackage[utf8]{inputenc} 
\usepackage[T1]{fontenc}    
\DeclareUnicodeCharacter{0301}{/}  
\DeclareUnicodeCharacter{2212}{-}  
\usepackage{ae,aecompl}
\usepackage{pgf,pgfarrows,pgfnodes,pgfautomata,pgfheaps}
\usepackage{graphicx}       
\usepackage{natbib}         
\usepackage{url}            
\usepackage{grffile}        
\usepackage{orcidlink}
\usepackage{mathtools}      
\usepackage{multirow}       
\usepackage{xspace}         

\usepackage{txfonts}
\usepackage{amsmath,amssymb,amsxtra,amsfonts}   

\usepackage{color}         
\definecolor{gold}{rgb}{1,0.80,0}
\definecolor{orange}{rgb}{1,0.5,0}
\definecolor{midgray}{gray}{0.3}
\definecolor{lblue}{rgb}{0,0.2,0.6}
\definecolor{dgreen}{rgb}{0.1,0.6,0.3}
\definecolor{purple}{rgb}{0.5019607843137255,0.0,0.5019607843137255}

\renewcommand\farcs{\mbox{$.\!^{\prime\prime}$}}    
\renewcommand{\arcdeg}{\mbox{$^{\circ}$}\xspace}             

\newcommand{\be}{\begin{equation}}
\newcommand{\ee}{\end{equation}}

\newcommand{\ba}{\begin{align}}
\newcommand{\ea}{\end{align}}

\newcommand{\defeq}{\vcentcolon=}

\newcommand{\Msun}{\ensuremath{M_\odot}\xspace}

\newcommand{\oh}{\ensuremath{12+\log({\rm O/H})}\xspace}

\newcommand{\K}{\ensuremath{\rm K}\xspace}

\newcommand{\Hunit}{\ensuremath{\rm km~s^{-1}~Mpc^{-1}}\xspace}
\newcommand{\Funit}{\ensuremath{\rm erg~s^{-1}~cm^{-2}}\xspace}

\newcommand{\Hb}{\textrm{H}\ensuremath{\beta}\xspace}
\newcommand{\Hg}{\textrm{H}\ensuremath{\gamma}\xspace}

\newcommand{\HII}{\textrm{H}\textsc{ii}\xspace}

\newcommand{\OII}{[\textrm{O}\textsc{ii}]\xspace}
\newcommand{\OIII}{[\textrm{O}\textsc{iii}]\xspace}

\newcommand{\NII}{[\textrm{N}\textsc{ii}]\xspace}

\newcommand{\NeIII}{[\textrm{Ne}\textsc{iii}]\xspace}

\def\U{\ensuremath{U_{360}}\xspace}     
\def\B{\ensuremath{B_{435}}\xspace}
\def\V{\ensuremath{V_{606}}\xspace}
\def\I{\ensuremath{I_{814}}\xspace}
\def\Y{\ensuremath{Y_{105}}\xspace}
\def\J{\ensuremath{J_{125}}\xspace}

\def\H{\ensuremath{H_{160}}\xspace}

\def\p{{\rm prior}}

\newcommand{\Om} {\ensuremath{\Omega_{\rm{m}}}\xspace}

\newcommand{\Ol} {\ensuremath{\Omega_{\Lambda}}\xspace}

\newcommand{\n}{\ensuremath{{\nu}\rm}\xspace}

\usepackage{float,longtable}
\usepackage{orcidlink}
\usepackage{booktabs} 
\usepackage{soul}
\usepackage{CJK}

\newcommand{\yy}[1]{{\color{black}{ #1}}} 

\begin{document}
\begin{CJK*}{UTF8}{gbsn}

\title{MAMMOTH-Grism: Revisiting the Mass-Metallicity Relation in Protocluster Environments at Cosmic Noon}

\correspondingauthor{Xin Wang, Chao-Wei Tsai}
\email{xwang@ucas.ac.cn, cwtsai@nao.cas.cn}


\author[0000-0002-0663-814X]{Yiming Yang}
\affil{School of Astronomy and Space Science, University of Chinese Academy of Sciences (UCAS), Beijing 100049, China}
\affil{National Astronomical Observatories, Chinese Academy of Sciences, Beijing 100101, China}

\author[0000-0002-9373-3865]{Xin Wang}
\affil{School of Astronomy and Space Science, University of Chinese Academy of Sciences (UCAS), Beijing 100049, China}
\affil{National Astronomical Observatories, Chinese Academy of Sciences, Beijing 100101, China}
\affil{Institute for Frontiers in Astronomy and Astrophysics, Beijing Normal University,  Beijing 102206, China}

\author[0000-0002-1336-5100]{Xianlong He}
\affil{School of Astronomy and Space Science, University of Chinese Academy of Sciences (UCAS), Beijing 100049, China}
\affiliation{Department of Physics and Astronomy, University of California, Los Angeles, 430 Portola Plaza, Los Angeles, CA 90095, USA}
\affiliation{Department of Astronomy, School of Physics and Technology, Wuhan University, Wuhan 430072, China}

\author[0000-0002-9390-9672]{Chao-Wei Tsai}
\affil{National Astronomical Observatories, Chinese Academy of Sciences, Beijing 100101, China}
\affil{Institute for Frontiers in Astronomy and Astrophysics, Beijing Normal University,  Beijing 102206, China}
\affil{School of Astronomy and Space Science, University of Chinese Academy of Sciences (UCAS), Beijing 100049, China}


\author[0000-0001-8467-6478]{Zheng Cai}
\affiliation{Department of Astronomy, Tsinghua University, Beijing 100084, China}

\author[0000-0001-5951-459X]{Zihao Li}
\affiliation{Cosmic Dawn Center (DAWN), Denmark}
\affiliation{Niels Bohr Institute, University of Copenhagen, Jagtvej 128, DK2200 Copenhagen N, Denmark}
\affiliation{Department of Astronomy, Tsinghua University, Beijing 100084, China}

\author[0000-0001-6919-1237]{Matthew A. Malkan}
\affiliation{Department of Physics and Astronomy, University of California, Los Angeles, 430 Portola Plaza, Los Angeles, CA 90095, USA}

\author[0000-0002-3264-819X]{Dong Dong Shi}
\affiliation{Center for Fundamental Physics, School of Mechanics $\&$ Optoelectronic Physics, Anhui University of Science and Technology, Huainan 232001, China}


\author[0000-0002-8630-6435]{Anahita Alavi}
\affiliation{Infrared Processing and Analysis Center, Caltech, 1200 E. California Blvd., Pasadena, CA 91125, USA}

\author[0000-0002-1620-0897]{Fuyan Bian}
\affiliation{European Southern Observatory, Alonso de Cordova 3107, Casilla 19001, Vitacura, Santiago 19, Chile}

\author[0000-0001-6482-3020]{James Colbert}
\affiliation{Infrared Processing and Analysis Center, Caltech, 1200 E. California Blvd., Pasadena, CA 91125, USA}

\author[0000-0003-3310-0131]{Xiaohui Fan}
\affiliation{Steward Observatory, University of Arizona, 933 North Cherry Ave., Tucson, AZ 85721, USA}

\author[0000-0002-6586-4446]{Alaina L. Henry}
\affiliation{Space Telescope Science Institute, 3700 San Martin Dr., Baltimore, MD, 21218, USA}

\author[0000-0002-7064-5424]{Harry I. Teplitz}
\affiliation{Infrared Processing and Analysis Center, Caltech, 1200 E. California Blvd., Pasadena, CA 91125, USA}

\author[0000-0003-3728-9912]{Xian~Zhong Zheng}
\affiliation{Tsung-Dao Lee Institute and State Key Laboratory of Dark Matter Physics, Shanghai Jiao Tong University, Shanghai 201210, China}



\begin{abstract}

We present one of the first measurements of the mass-metallicity relation (MZR) in multiple massive protoclusters at cosmic noon, using Hubble Space Telescope (HST) G141 slitless spectroscopy from the MAMMOTH-Grism survey. We identify 63 protocluster member galaxies across three overdense structures at $z = 2\text{--}3$ with robust detections of [\ion{O}{3}], H$\beta$, and [\ion{O}{2}] emission. The sample spans gas-phase metallicities of $12 + \log(\text{O/H}) = 8.2\text{--}8.6$, dust-corrected H$\beta$-based star formation rates (SFRs) of $10$-$250\,M_\odot\,\text{yr}^{-1}$, and stellar masses of $M_\ast \sim 10^{9.4}$-$10^{10.5}\,M_\odot$, derived via spectral energy distribution fitting using deep HST and ground-based photometry. 
We stack spectra in five $M_\ast$ bins to obtain average metallicities and SFRs.
Relative to field galaxies at similar redshifts, protocluster members show elevated SFRs at $M_\ast < 10^{10.25}\,M_\odot$ and a systematically shallower MZR: $12 + \log(\text{O/H}) = (6.96 \pm 0.13) + (0.143 \pm 0.017) \times \log(M_{\ast}/M_{\odot})$. 
We detect a mass-dependent environmental offset: massive protocluster galaxies are metal-poor compared to field counterparts of similar mass, whereas lower-mass systems exhibit comparable or mildly enhanced metallicities. This trend is consistent with a scenario where cold-mode accretion dilutes the interstellar medium (ISM) across the full mass range, while efficient recycling of feedback-driven outflows preferentially enriches the ISM in low-mass galaxies.
Finally, we assess the dependence of metallicity offsets on local overdensity and find no significant trend, likely reflecting the survey's bias toward protocluster cores.

\end{abstract}

\keywords{Protoclusters --- Galaxy evolution --- Galaxy abundances --- Galaxy formation --- High-redshift galaxies} 


\section{Introduction} \label{sec:intro}
The gas-phase metallicity (hereafter referred to as metallicity) is a crucial observational diagnostic parameter in the study of galaxy formation and evolution. This parameter can be influenced by many processes, including star formation, gas inflow, gas outflow \citep{1985ApJ...290..154L, 2015MNRAS.452.4361K, 2016MNRAS.457.2605C, 2018MNRAS.474.1143L,2021ApJ...910..137W, 2023MNRAS.518.5500B,2025arXiv250612129L}, and AGN feedback \citep{2013MNRAS.435.2931H,2017MNRAS.468..751E, 2019SSRv..215....5W,2020ApJ...904....8C,2023ApJ...958..147P,2024MNRAS.52711043Y}. The environment plays a significant role in these processes throughout cosmic time \citep{2014ARA&A..52..291C,2018A&A...617A..53A,2024A&A...691A.341T,2024MNRAS.528.4393G}, with emerging evidence suggesting the onset of environmental effects as early as redshift $z \gtrsim 5$ \citep{Helton_24, Morishita_25, Li_25}.
During the cosmic noon ($z\sim2-3$), a critical epoch when galaxies experienced their most vigorous growth, we expect to observe more significant environmental effects on those processes. 
\yy{In this active epoch, an ideal laboratory for developing our understanding of the physical phenomena associated with this impact is galaxy protoclusters, which are the progenitors of today's massive galaxy clusters \citep{2013ApJ...779..127C, 2015MNRAS.452.2528M, 2016A&ARv..24...14O, 2017ApJ...844L..23C}.
Recent studies have explored different aspects of galaxy formation and evolution in protocluster environments,
including the star formation rate (SFR) \citep{2005ApJ...621..201C,2011MNRAS.415.2993H, 2017ApJ...844L..23C,2017MNRAS.464..876H,2024A&A...683A..57R,2024MNRAS.531.2335D}, quenching \citep{2016ApJ...817....9K, 2017ApJ...847..134K, 2017MNRAS.472.3512T, 2024MNRAS.527.8598E, 2025ApJ...990L..24P}, active galactic nuclei (AGN) activity \citep{2024A&A...683A..57R,2024A&A...689A.130V,2025A&A...694A.165T,2025arXiv251012393D}, cold gas content \citep{2017A&A...608A..48D,2025A&A...696A.236P,2025A&A...701A.234Z,2025arXiv251012393D} 
and many other aspects \citep{kulasMASSMETALLICITYRELATION2013, Shimakawa2015, 2015ApJ...801..132V,2018shimakawa, Sattari2021,chartabMOSDEFSurveyEnvironmental2021,Wang_2022,2022ApJ...929L...8L,2023MNRAS.523.2422L, perez-martinezEnhancedStarFormation2024,2025arXiv250504212Z,2025ApJ...990..225B}.
One of the key diagnostics for understanding the interplay between galaxy properties and their environments is the mass - metallicity relation (MZR). The MZR defines a well-established correlation between a galaxy's stellar mass and its metallicity, observed consistently in both the local and distant universe \citep{2004ApJ...613..898T,2006ApJ...644..813E,2013ApJ...765..140A, maiolino2019, Sanders2021,Henry.2021, calabr2022}. }
Furthermore, the influence of environment on galaxy metallicity has been extensively studied at low redshift ($z\sim0$), revealing that galaxies in denser regions, such as clusters and groups, 
tend to exhibit higher metallicities compared to those in less dense, field regions \citep{2006MNRAS.373..469B}.
However, the relationship between environment and metallicity becomes more complex and remains under active investigation at cosmic noon, with conflicting evidence of the existence of any environmental effects \citep{kulasMASSMETALLICITYRELATION2013,2015ApJ...802L..26K, Shimakawa2015, 2015ApJ...801..132V, Sattari2021,chartabMOSDEFSurveyEnvironmental2021,Wang_2022, 2023MNRAS.518.1707P, perez-martinezEnhancedStarFormation2024}.

\cite{MAMMOTH2016} first confirmed the existence of the extremely massive overdensities at $z\sim2-3$ traced by groups of Coherently Strong Intergalactic Ly$\rm \alpha$ Absorption (CoSLA) using a technique referred to as MApping the Most Massive Overdensity Through Hydrogen (MAMMOTH).
MAMMOTH utilizes the largest library of quasar spectra, SDSS-III Baryon Oscillations Spectroscopic Survey (BOSS), to locate strong HI absorption and identify the most massive protocluster candidates \citep[e.g.,][]{liang2021}. 
Using the MAMMOTH technique, the protoclusters BOSS1244 ($z=2.24\pm0.02$), BOSS1441($z=2.32\pm0.02$) and BOSS1542 ($z=2.24\pm0.02$) were selected from the early data release of SDSS-III BOSS and confirmed by the follow-up observations \citep{MAMMOTH2017,MAMMOTH2017B,2021MNRAS.500.4354Z,2021ApJ...915...32S}. 
These protoclusters were selected based on their extreme overdensities and their potential to evolve into structures more massive than the Coma cluster, making them excellent candidates for studying the effects of environment on galaxy formation and evolution.

In this paper, we aim to disentangle environmental effects from the intrinsic properties of galaxies and to understand how these factors influence the MZR and the star formation rate (SFR).
Extending our previous work on BOSS1244 \citep{Wang_2022}, we incorporated improved photometric measurements and included two additional protoclusters (BOSS1441 and BOSS1542). With the expanded sample, we presented one of the first measurements of the MZR in multiple massive protoclusters at cosmic noon.
We employed the deep Hubble Space Telescope (HST) Wide-Field Camera 3 (WFC3) grism spectroscopy and multi-wavelength broadband imaging to derive the stellar masses, SFRs, and metallicities of these galaxies.
In addition, we analyzed the relationship between the metallicity and the galaxy overdensity $\delta_g$ in these protoclusters. 

This paper is structured as follows. In Section \ref{obs}, we describe the observations and the data reduction process. 
In Section \ref{measure}, we outline the methods used to derive stellar masses, SFRs, and metallicities for our sample galaxies, along with the stacking procedure and AGN removal.
In Section \ref{rsl}, we present the MZR of galaxies in overdense environments, the metallicity offset compared to field galaxies, and the relationship between galaxy
overdensity and metallicity offset.
In Section \ref{discuss}, we interpret our findings in the context of environmental effects on galaxy evolution, focusing on gas accretion and feedback processes.
Finally, we summarize our conclusions and their implications for galaxy formation in Section \ref{con}.
Throughout this paper, we adopt the AB magnitude system and the standard concordance cosmology ($\Om=0.3, \Ol=0.7$, $H_0=70\,\Hunit$).  
The metallic lines are denoted in the following manner, if presented without wavelength:

$\OIII\lambda$5008$\defeq$\OIII,
$\OII\lambda\lambda$3727,3730$\defeq$\OII,

$\NeIII \lambda $3869$\defeq$\NeIII, 
and $\NII{\lambda}$6585$\defeq$\NII.

\section{Observations} \label{obs}

\yy{ We have conducted an HST Cycle-28 medium program, the MAMMOTH-Grism survey (GO-16276, P.I.: X. Wang), 
a slitless spectroscopic survey designed to obtain deep WFC3/G141 grism and F125W imaging data in three of the most massive galaxy protoclusters at $z = 2-3$ (see Figure \ref{fig:protoclusters} in Appendix).
Each protocluster core was mapped with five WFC3/IR pointings, yielding a total effective area of $\sim$ 20 arcmin$^2$ per field.
For each pointing, we obtained exposures at three position angles (i.e., 10-15 degrees) to minimize spectral overlap and confusion.
The G141 has a spectral resolution of R = 130 and covers a wavelength range of 1.1--1.7 $\mu$m, allowing us to observe key rest-frame optical emission lines such as [\ion{O}{3}]$\lambda\lambda$4960,5008, H$\beta$, and [\ion{O}{2}]$\lambda\lambda$3727,3730 for galaxies at $z \sim 2-3$.
The total integration times are about 30 ks for G141 spectroscopy and about 9 ks for F125W imaging per field. }

\yy{We also leveraged the existing multi-wavelength imaging data available in these fields, including HST/WFC3 F160W, LBT/LBC, CFHT/WIRCam, and KPNO/MOSAIC imaging data,
the details of which are summarized in Table \ref{tab:photometry}.}
These images were processed using the AstroDrizzle software and aligned with the GAIA DR2 astrometric frame.

\begin{deluxetable*}{lccccc}
\tablecaption{Summary of the photometric datasets used in this work.}
\tabletypesize{\scriptsize}
\tablecolumns{6}
\tablewidth{0pt}
\tablehead{
    \colhead{Telescope/Instrument} & 
    \colhead{Filter} & 
    \colhead{Integration Time} & 
    \colhead{Angular Resolution (arcsec)} & 
    \colhead{5$\sigma$ Limiting Magnitude (AB)} & 
    \colhead{Protocluster}
}
\startdata
\multicolumn{6}{c}{\textbf{Space-based Observations}} \\
\noalign{\smallskip}
HST/WFC3 & F125W & $\sim$1.8 ks  & 0\farcs13 & 24.95 & All three protoclusters \\
HST/WFC3 & F160W & $\sim$2.6 ks  & 0\farcs15 & 24.97 & All three protoclusters \\
\noalign{\smallskip}\hline\noalign{\smallskip}
\multicolumn{6}{c}{\textbf{Ground-based Observations}} \\
\noalign{\smallskip}
LBT/LBC & $U_{\rm spec}$ & 4.7 hr, 6.0 hr, 4.8 hr & 0\farcs96, 0\farcs60, 0\farcs80 & 26.7, 26.5, 26.4 & BOSS1244, BOSS1441, BOSS1542 \\
LBT/LBC & $V_{\rm BESSEL}$ & 2.5 hr & 0\farcs60 & 26.4 & BOSS1441 \\
LBT/LBC & Sloan $z$ &  4.0 hr, 9.2 hr & 0\farcs81, 0\farcs90 & 25.1, 26.2 & BOSS1244, BOSS1542 \\
CFHT/WIRCam & $K_s$ & 5.0 hr, 5.2 hr & 0\farcs78 & 23.3, 23.2 & BOSS1244, BOSS1542 \\
KPNO-4m/MOSAIC & $Bw$ & 3.0 hr & 1\farcs32 & 25.9 & BOSS1441 \\
\enddata
\tablecomments{
Integration times for the two HST/WFC3 filters are given per protocluster field. 
The angular resolution of ground-based imaging refers to the median seeing values.
The 5$\sigma$ limiting magnitudes are measured within a circular aperture of 1.0 arcsec diameter for the HST data and 2.0 arcsec for the ground-based data.
}
\label{tab:photometry}
\end{deluxetable*}


We utilized the GRIZLI software \citep{2019ascl.soft05001B} to reduce
 the WFC3/G141 data. This software operates through a series of steps
  to analyze the paired direct imaging and grism exposures.
\begin{enumerate}
    \item Pre-processing of the raw WFC3 exposures, which includes tasks 
    such as relative/absolute astrometric alignment, flat-fielding, and 
    variable sky background subtraction due to He I atmospheric airglow.
    \item Forward-modeling of the full field of view (FoV) grism exposures
     at the visit level.
    \item Redshift fitting via spectral template synthesis, which involves
     fitting linear combinations of spectral templates to the observed optimally
      extracted 1D grism spectra to determine the best-fit grism redshifts.
      \yy{These grism redshifts were derived from simultaneous fitting of several emission lines。 }
    \item Refinement of the full-FoV grism model.
    \item Extraction of 1D/2D grism spectra and emission-line maps for individual extended objects.
\end{enumerate}

\section{Measurements}\label{measure}

\subsection{Sample Selection} 
\begin{figure}[htb]
    \center
    \includegraphics[width=1\linewidth]{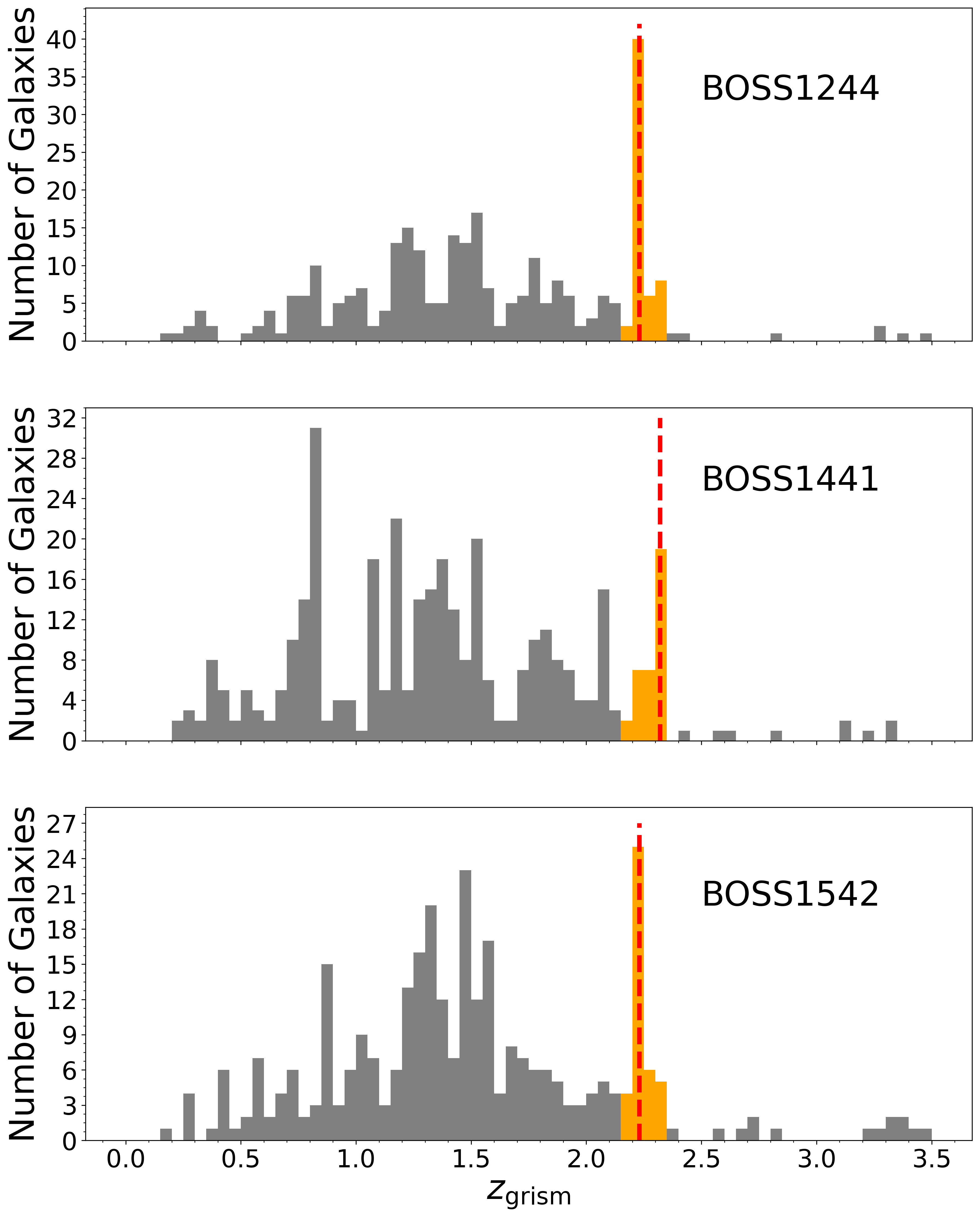}
    \caption{Histograms showing the distribution of grism redshifts ($z_{\rm grism}$) for galaxies along the line of sight toward the core regions of the three protoclusters (BOSS1244, BOSS1441, and BOSS1542). 
    Only galaxies with secure grism redshifts, as defined in Section 3.1, were included. \yy{The top panel corresponds to BOSS1244, which is adapted from \cite{Wang_2022}.}
    The red dashed line indicates the spectroscopic redshift ($z_{\rm spec}$) of BOSS1244 ($z=2.24$), BOSS1441 ($z=2.32$) and BOSS1542 ($z=2.24$),
    measured from ground-based multi-object spectroscopy with higher wavelength resolution than that of WFC3 G141 \citep{MAMMOTH2017B,2021ApJ...915...32S}.    
    The orange-filled bars highlight galaxies within the redshift range $z \sim [2.15, 2.35]$, where the protocluster member galaxies were selected. }
    \label{fig:redshift.png}
\end{figure}

The grism redshift fit for each source was considered secure only 
if it met certain goodness-of-fit criteria, including a reduced 
chi-squared value less than two, the width of the redshift posterior 
($\Delta z_{\rm posterior} / (1+z_{ \rm peak})$) being less than 0.005 and the 
Bayesian information criterion being more than 30. 

In the redshift deriving procedure, we
performed a fit for the intrinsic nebular emission lines to our samples with 1D Gaussian profiles, 
resulting in the acquisition of flux measurements of \OII, H$\delta$, \Hb, \Hg, \NeIII, and \OIII.
Then, to secure accurate
metallicity estimates, we selected sources with SNRs more than three in \OIII and \OII.
\yy{The sample selection procedures resulted in a total of 77 member galaxies: 18 from BOSS1441, 20 from BOSS1542, and 39 from BOSS1244 (which were previously identified in Wang+22) with secure grism redshifts in the
range of $z \sim [2.15, 2.35]$ and prominent nebular emission features
suitable for our analyses (see Figure \ref{fig:redshift.png}).} 
\yy{To double-check, we visually review the 1D spectrum to exclude sources that may be contaminated by nearby objects, which caused strong continua and abrupt flux jumps.
There are 6 sources removed in this step.} After removing active galactic nuclei (AGNs; see Section \ref{AGN}), a sample of 63 galaxies was retained for the final analysis.

\subsection{Stellar Mass}
We performed aperture photometry on the HST images using SExtractor \citep{1996A&AS..117..393B} in  dual-image mode,
with the stacked F125W and F160W mosaics serving as the detection image to increase the depth. 

To account for the resolution differences between HST and ground-based images, we used the T-PHOT software \citep{Merlin.2016}, which fits the low-resolution images using templates derived from high-resolution images. This approach enables accurate photometric measurements across a broad wavelength range through a multi-step procedure to model low-resolution images based on high-resolution templates and perform the photometric calibration.
\yy{Here, we used the same grism and imaging data of BOSS1244 as in our previous study, while the datasets for BOSS1441 and BOSS1542 are newly incorporated. We also improved the photometric analysis by employing newly constructed effective PSFs (ePSFs) in the T-PHOT fitting process to enhance the flux accuracy. The use of these new ePSFs reduces residuals in the photometric fitting compared with our previous analysis, resulting in more reliable stellar mass measurements. To ensure uniformity across the sample, the new ePSFs were applied to all three protoclusters, including BOSS1244.
}

\begin{enumerate}
    \item Create cutouts from the high-resolution images using SExtractor outputs.
    \item Match the PSFs of the high-resolution images to those of the low-resolution images using Fourier convolution, resulting in a kernel with the same pixel scale as the high-resolution image.
    \item Create low-resolution templates by convolving the high-resolution cutouts with the kernel obtained in the previous step.
    \item Fit the templates to the low-resolution images to derive the photometric measurements.
\end{enumerate}
By minimizing the difference between the observed images and the templates, we obtain accurate photometric measurements from the ground-based images that are consistent with the isophotal fluxes of the HST images \citep{2007PASP..119.1325L}.
These measurements allow us to robustly estimate the stellar masses of the sample galaxies via spectral energy distribution (SED) fitting. 
We use the BAGPIPES software \citep{2018MNRAS.480.4379C}
to fit the BC03 \citep{2003MNRAS.344.1000B} 
SED models to the photometry, assuming an 
initial mass function from \cite{2003PASP..115..763C} 
an extinction law from \cite{Av2000C} \yy{with prior listed as Table \ref{tab:bagpipes_priors}}.
We adopted an exponentially declining star formation history (SFH) model in the fitting process.
The best fit grism redshifts were adopted with a conservative uncertainty of $\Delta z/(1+z) \simeq 0.003$ 
\citep{2016ApJS..225...27M}. We also found that the actual errors 
estimated by GRIZLI were smaller than this value \citep{Wang_2022}, as our protoclusters lie at redshifts where the G141 grism is particularly efficient and accurate for redshift measurements.
Given that the selected galaxies exhibit strong emission lines, we included nebular emission in the SED fitting. 
\yy{We include some of our SED fitting cases in the Appendix FIgure \ref{fig:SED_fitting_cases}.}
\begin{deluxetable}{lcccccccccccc}[htb!]
    \tablecolumns{3}
    \tablewidth{4pt}
    \tablecaption{Priors Used in the Bagpipes Fitting Analysis} 
\tablehead{
    \colhead{Parameter} &
    \colhead{Range} &
    \colhead{Prior Description} 
}
\startdata
    $Z$       &  0 - 2.5 dex            & Gas-phase metallicity\\
    $A_{\rm V}$ & 0 - 3      & Dust extinction  \\
    SFH & $\tau $ = 0.01 - 10 Gyr & Star formation history \\
\enddata
    \tablecomments{SFH follows an exponentially declining model. The time since star formation began is allowed to vary from 0.01 Gyr up to the age of the Universe at the galaxy's redshift, while the  e-folding  timescale ($\tau $) varies from 0.01 to 10 Gyr.
    }

\label{tab:bagpipes_priors}
\end{deluxetable}

\subsection{The Gas-Phase Metallicity}
We used the observed emission line fluxes (e.g, \OII , \OIII, \Hg, \Hb) and their uncertainties to obtain three parameters: metallicity, nebular dust extinction, and the dereddened $\rm H\beta$ line flux
$(12 + \log(\text{O/H})$,  $A_{\rm V}$,  $f_{\rm H\beta})$. Building upon our previous works \citep{wang2017,wang2019,wang2020b,Wang_2022}, we applied a forward-modeling Bayesian inference method.
\yy{The key advantage of this approach is that, rather than calculating observed emission line flux ratios, it directly forward-models the observed line fluxes.}
This approach is superior to conventional methods as it marginalizes over faint, low-SNR lines \citep[see][Figure 6]{wang2020b}. 
We adopted the extinction law \cite{Av2000C} to correct for dust attenuation
and derived $A_{\rm V}$ from the parameter sampling. \yy{ In practice, the extinction is corrected during the parameter-sampling process, as each $A_V$ 
is drawn from the 3D parameter space $(12+\log{\rm O/H}, A_V, f_{\rm H\beta})$ via Markov Chain Monte Carlo (MCMC) algorithm.}
The likelihood is defined as:
\begin{equation}
    L \propto \exp\left(-\frac{1}{2} \sum\frac{f_{EL_i}-R_i\cdot f^2_{\rm H\beta}}{\sigma ^2_{EL_i} + f^2_{\rm H\beta} \cdot \sigma ^2_{R_i}}\right) 
\end{equation}
where $R_{i}$ denotes the expected line flux ratio, either the Balmer decrement ($\mathrm{H}\beta/\mathrm{H}\gamma = 0.47$) or a metallicity diagnostic such as $\OIII/\mathrm{H}\beta$. The observed emission-line fluxes and their uncertainties are denoted by
 $f_{{EL}}$ and $\sigma_{{EL}}$, respectively. We used the emcee sampler \citep{emcee} for posterior sampling during the calculation.
\yy{In this work, we adopt the empirical strong-line calibrations of \citet{2018ApJ...859..175B} (hereafter B18), derived from local analogs of high-z galaxies selected by their BPT diagram positions.
    The coefficients for these strong line calibrations are given in Table \ref{tab:coef}.} \yy{We stress that metallicities shown below are on the empirical calibration of B18 applied consistently across our sample; absolute values may shift under alternative calibrations and therefore inter-study comparisons should be interpreted in a relative sense.}

\begin{deluxetable}{lcccccccccccc}
    \tablecolumns{5}
    \tablewidth{0pt}
    \tablecaption{Coefficients for Emission Line Ratio Diagnostics}
\tablehead{
    \colhead{$\log(R)$} &
    \colhead{$c_0$} &
    \colhead{$c_1$} &
    \colhead{$c_2$} &
    \colhead{$c_3$} 
}
\startdata
    \multicolumn{5}{c}{Strong line calibrations of \citet{2018ApJ...859..175B}}   \\
    \noalign{\smallskip}
    \OIII$\lambda\lambda$4960,5008/\Hb   &   43.9836    &   -21.6211   &   3.4277  &   -0.1747   \\
    \OII/\Hb    &   78.9068    &   -45.2533   &   7.4311  &   -0.3758   \\
    \noalign{\smallskip}\hline\noalign{\smallskip}
    \multicolumn{5}{c}{Balmer decrement}   \\
    \noalign{\smallskip}
    \Hg/\Hb   &   -0.3279    &   \nodata   &   \nodata  &   \nodata \\
\enddata
    \tablecomments{The empirical flux ratios are computed in the polynomial functional form of $\log{R} = \sum_{i} c_i\cdot x^i$, where $x=\oh$.
    }

\label{tab:coef}
\end{deluxetable}

\yy{We note that our selection requiring S/N $>$ 3 in both [O III] and [O II] may introduce biases in the resulting metallicity distribution. Because weaker [O III] emission is typically associated with higher gas metallicity and lower ISM temperature, such a criterion could preferentially exclude metal-rich systems, particularly among massive galaxies. Consequently, our sample may be slightly skewed toward lower-metallicity sources compared to studies using [N II]/H$\alpha$-based diagnostics, which are more sensitive to the high-metallicity regime. }
\subsection{Star Formation Rate }
We used the Balmer line luminosities to estimate the instantaneous SFR of our sample galaxies. 
\yy{Using the forward-modeling Bayesian method, we can conduct constraints on the dereddened $\rm H\beta$ flux.
We note that the $\rm H\beta$ emission line we used as an SFR calibrator, may be affected by underlying stellar absorption from older populations, particularly in massive galaxies.
Consequently, while our forward modeling yields tight constraints, these should be interpreted as lower limits on the true uncertainties, which include systematic effects from stellar absorption.}
Assuming the  \cite{1998ARA&A..36..189K} calibration and the Balmer decrement ratio of $\rm H\alpha/H\beta = 2.86$ from the case B recombination for typical \HII regions, 
we can obtain the SFR of our sample galaxies as follows, which is suitable for the \cite{2003PASP..115..763C} initial mass function.

\begin{equation}
    \text{SFR} = 1.33 \times 10^{-41} \frac{L(\text{H}\beta)}{\rm [erg\ s ^{-1}]} [M_\odot\ yr ^{-1}] 
\end{equation}
The derived $M_\star$ and SFR values of our galaxy samples are listed in Table \ref{tab:indvd} in the Appendix. In Figure \ref{SFR}, we present the SFR - $M_\ast$ relation of the protoclusters members in gray.
\begin{figure}[htb]
    \centering
    \includegraphics[width=1\linewidth]{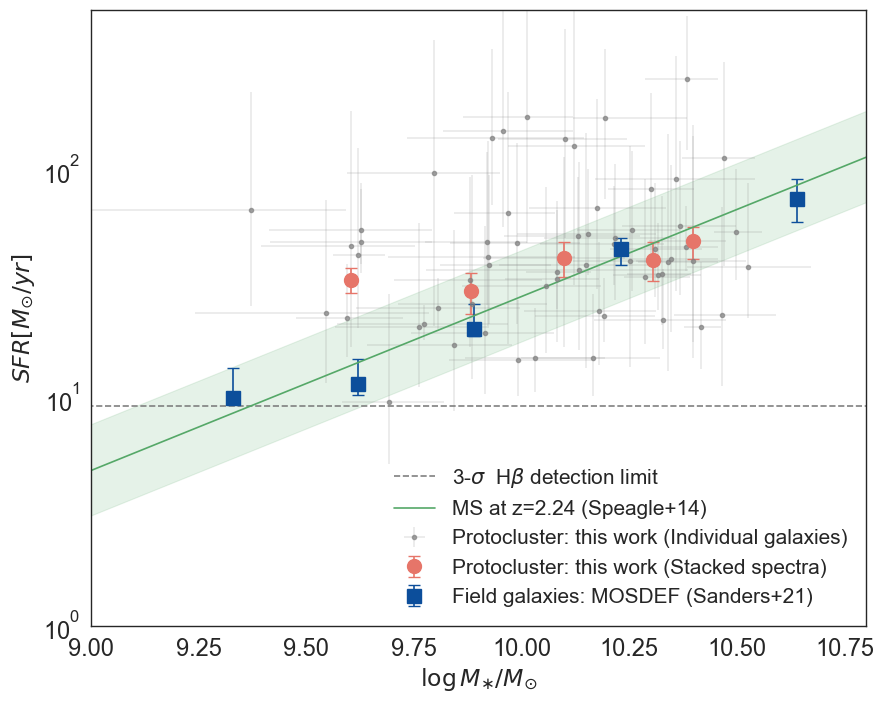}
    \caption{SFR versus $M_\ast$ of individual protocluster galaxies (gray dots) and the stacked sample medians (orange dots). For comparison, we include the stacked field galaxy sample from  \citet{Sanders2021} shown in blue. The green shaded region denotes the star-forming main sequence at $z=2.24$ derived by \citet{2014ApJS..214...15S}, with a 0.2 dex scatter.
    }
    \label{SFR}
\end{figure}

\subsection{AGN Identification and Removal}\label{AGN}
\begin{figure}[htb]
    \centering
    \includegraphics[width=1\linewidth]{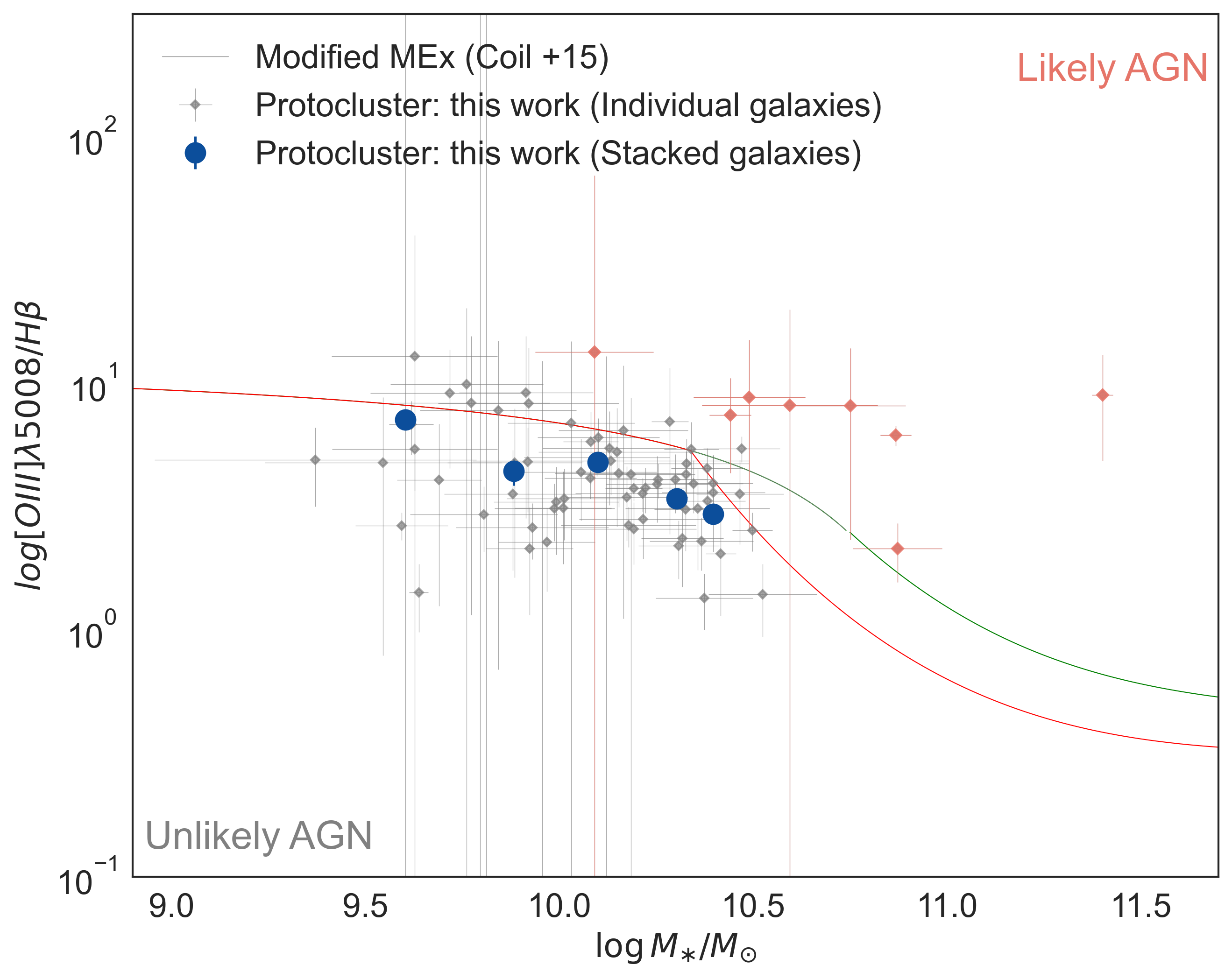}
    \caption{MEx diagram for the galaxies in the three protoclusters. 
    Individual galaxies are shown in gray, while sources classified as AGNs are highlighted in red. 
    The green and red lines represent the empirical separation boundaries between star-forming galaxies and AGNs, 
    following \cite{2015ApJ...801...35C}. 
    \yy{Accounting for the uncertainties in the measurements and in the applying of MEx diagram, we adopt an tolerance of 0.15 dex in the classification.}
    The stacked subsamples used in the metallicity analysis are shown in blue.} 
    \label{BPT}
\end{figure}

We identified and removed AGNs using the mass-excitation (MEx) diagram to distinguish
between AGNs and \HII regions in emission-line galaxies. 
The well-known MEx diagram, proposed by \cite{2011ApJ...736..104J, 2014ApJ...788...88J} and modified by \cite{2015ApJ...801...35C}, 
utilizes ($M_\ast$) as a proxy for the \NII /H$\alpha$ ratio to 
differentiate between AGNs and \HII regions. 
By applying this modified version, we removed AGN contamination from our galaxy sample, ensuring that the galaxies selected for stacking were primarily photoionized by massive stars in \HII 
regions rather than by AGNs. 
As shown in Figure \ref{BPT}, eight sources classified as AGNs by the MEx diagram were excluded from the stacking analysis described below.
\yy{
At z $\sim 2$, there is a high uncertainty using the MEx diagram (showing after) to identify AGNs, and the low SNR of individual galaxies adds to this uncertainty. 
Some of our sources in Figure \ref{BPT} have large error bars in y-axis, in the result of the low SNR of \Hb line.
Therefore, we exclude the AGNs using relatively relaxed (0.15 dex more tolerance) conditions. 
If we remove these sources at low mass bins, we will possibly include significant systematic errors here as we have very few low mass samples after removal.
}

\subsection{Spectral Stacking}\label{stacking}
In this subsection, we stacked the grism spectra of our sample galaxies to enhance the SNRs of the emission lines.
From the previous procedures, we found that the $M_\star$ of our galaxies range from $10^{9.3} $ to $10^{10.5} M_{\odot}$. To investigate potential mass-dependent trends, we divided the samples into five mass bins (see Table \ref{tab:stack}). 
We then utilized the following stacking procedures, similar to those in  \cite{Henry.2021}.
\begin{enumerate}
    \item Subtract continuum models constructed by the GRIZLI software from the observed grism spectra, combined from multiple orients. 
    \item  Normalize the continuum-subtracted spectrum of 
    each object by the measured [\OIII] flux, to avoid excessive weighting toward objects with stronger line fluxes.
    \item Deredden each normalized spectrum to its rest frame on a common wavelength grid.
    \item Take the median value of the normalized fluxes at each wavelength grid.
    \item Re-create the stacked spectra
     1000 times with bootstrapping replacement, and adopt the standard 
     deviation as the measurement uncertainty.
\end {enumerate}

\begin{figure*}[!htb]
    \centering
    \includegraphics[width=1\linewidth]{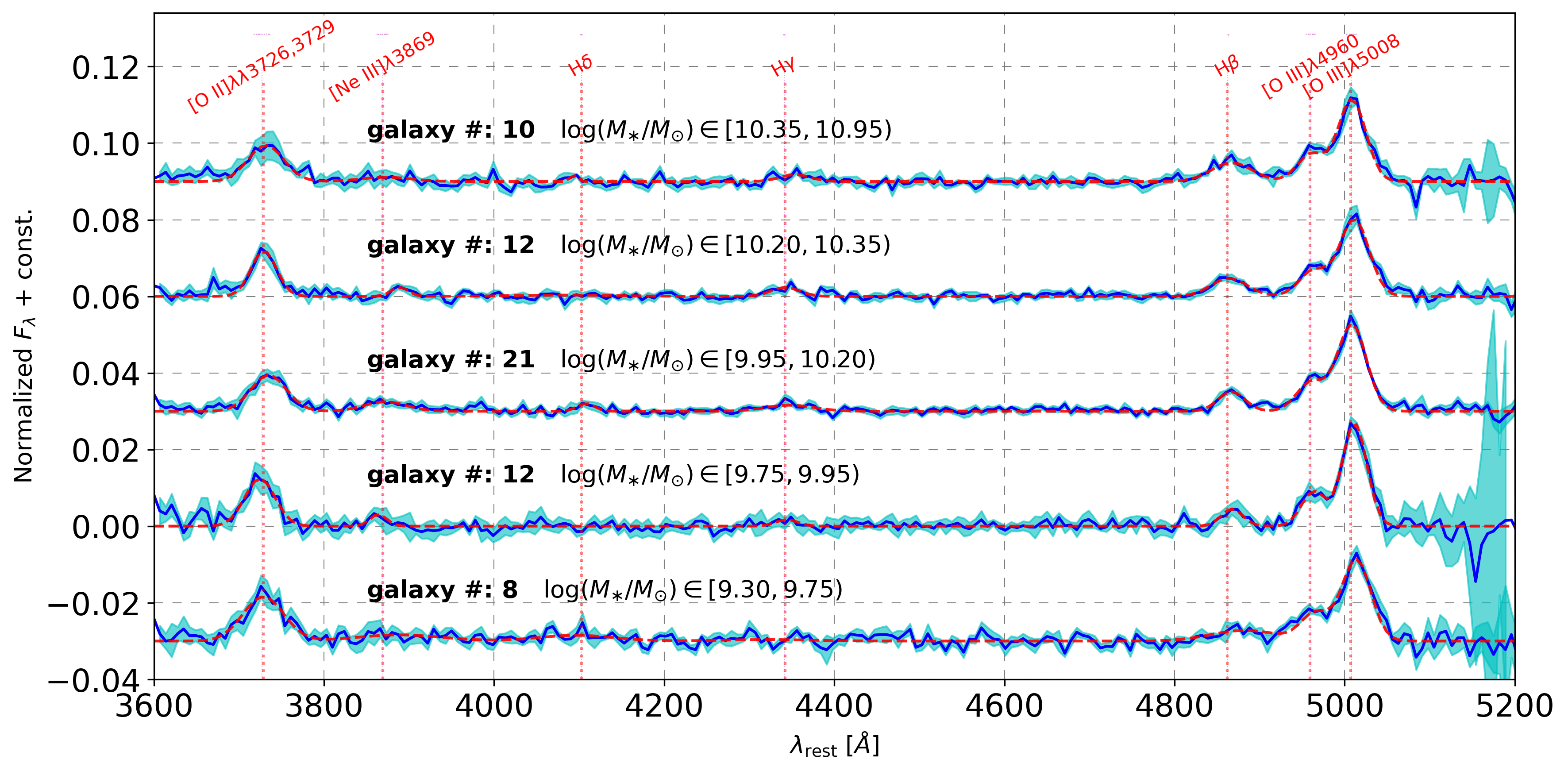}
    \caption{Stacked HST/WFC3 G141 grism spectra for galaxies in five stellar mass bins, divided as described in Table \ref{tab:stack}.
    The stacked spectra are shown in blue, with the 1$\sigma$ uncertainty envelopes indicated in light blue. 
    Red dashed lines show the best-fit Gaussian models to the emission lines derived using the LMFIT software \citep{lmfit}, and the red vertical dotted lines indicate the positions of the emission lines we used.
    }
    \label{5bins}
\end{figure*}

Next, we fit multiple Gaussian profiles to the stacked spectra using nonlinear least-squares minimization with LMFIT software \citep{lmfit} to derive line flux ratios within each mass bin. The Gaussian profiles were centered at the rest-frame wavelengths of emission lines (\OII. \NeIII, $\rm H\delta$, \Hg, \Hb, and $\OIII\lambda\lambda$ 4960, 5008 doublet). Here we fixed the amplitude ratio of the \OIII doublet to 1:2.98. following \citet{2000MNRAS.312..813S}.
The resulting stacked spectra and best fit emission line models (red dashed curves) are shown in Figure \ref{5bins}, and the line flux ratios are summarized in Table \ref{tab:stack}.
\yy{ For each stellar-mass bin, we derived the metallicity using the same Bayesian method as described in Section 3.3,
based on the emission line fluxes obtained from the stacked spectra. The stellar masses and SFRs of the stacked samples are calculated as the medians of the individual galaxies within each bin,
with their uncertainties estimated using bootstrapping resampling.
The derived metallicities, stellar masses, and SFRs of the stacked samples are listed in Table \ref{tab:stack}.
We need to stress here that as the SFRs of stacked sample shown in Figure \ref{SFR} are the medians of the individual galaxy SFRs within each stellar-mass bin, 
the error bars represent the 1$\sigma$ dispersion of those individual SFRs and therefore quantify the intrinsic scatter of the population rather than the measurement uncertainty propagated from the stacked spectra. }

{
\tabletypesize{\scriptsize}
\tabcolsep=2pt
\begin{deluxetable*}{lccccccccccc}
    \tablecolumns{12}
    \tablewidth{0pt}
    \tablecaption{Measured Properties of the Stacked Spectra}
\tablehead{
    \colhead{mass bin} &
    \colhead{$N_{\rm gal}$} &
    \colhead{log($M_{\ast}/M_{\odot}$)} &
    \colhead{$M_{\ast}^{\rm med}$}  &
    \colhead{SFR}   &
    \colhead{\OIII/\Hb}  &
    \colhead{\OII/\Hb}  &
    \colhead{\Hg/\Hb}  &
    \colhead{\OIII/\OII}  &
    \colhead{$f_{\OIII}$}  &
    \colhead{\oh}  &
    \colhead{$\Delta~\mathrm{log(O/H)_{cluster-field}}$}  \\\\
    & & range & [$M_{\odot}$] & [$M_{\odot}$/yr] &  &  &  &  & [$10^{-17}$\Funit]  &  \multicolumn{2}{c}{using B18 calibrations}  \\ 
}
\startdata
    1 & 10 &[10.35, 10.95) & 10$^{10.40}$ & $48.4\pm7.8$  & $2.96\pm0.23$ & $1.69\pm0.21$ & $0.26\pm0.11$ & $1.76\pm0.16$ & $9.47\pm1.13$ & $8.44_{-0.04}^{+0.05}$  & $-0.18\pm0.08$  \\
    2    &  12   & [10.20, 10.35) & 10$^{10.30}$ & $40.2\pm7.8$ & $3.42\pm0.30$ & $1.76\pm0.20$ & $0.38\pm0.06$ & $1.97\pm0.12$ & $10.57\pm1.89$  & $8.45_{-0.05}^{+0.05}$  & $-0.14\pm0.08$  \\
    3 & 21 & [9.95, 10.20) &  10$^{10.10}$ & $41.0\pm7.1$  & $4.80\pm0.43$ & $2.46\pm0.32$ & $0.57\pm0.21$ & $1.95\pm0.15$ & $10.60\pm1.73$  & $8.39_{-0.06}^{+0.05}$  & $-0.12\pm0.08$  \\
    4   &  12       & [9.75, 9.95) &  10$^{9.88}$ & $29.3\pm6.0$  & $4.41\pm0.55$ & $2.53\pm0.47$ & $0.17\pm0.31$ & $1.30\pm0.14$ & $9.03\pm1.82$  & $8.40_{-0.08}^{+0.07}$  & $-0.05\pm0.10$  \\
    5    &  8       & [9.30, 9.75) &  10$^{9.60}$ & $32.9\pm4.2$  &$6.80\pm0.20$ & $3.87\pm0.11$ & $0.38\pm0.22$ & $1.75\pm0.18$ & $7.50\pm1.81$  & $8.33_{-0.06}^{+0.05}$  & $  +0.01\pm0.08$ \\
\enddata
    \tablecomments{Physical and spectral properties of stacked galaxy samples in five stellar mass bins. 
    The multiple emission line flux ratios are measured from the stacked spectra shown in Figure ~\ref{5bins}. 
    The stellar masses, $f_{\OIII}$, and SFR results refer to the median value of galaxies within each mass bin, 
    with 1$\sigma$ uncertainty represented by the standard deviation.
    \yy{Stacked spectra are used solely for emission-line ratio measurements and metallicity inference via the forward-modelling procedure and do not enter into the calculation of the median SFRs. Consequently, a formally high S/N on a binned SFR (i.e., the low-mass bin where H$\beta$ and H$\gamma$ are weak or undetected in the stack) indicates a small spread in individual SFRs within the bin, not a high-significance detection of recombination lines in the stack.}
    }
\label{tab:stack}
\end{deluxetable*}
}

\section{Results}\label{rsl}
\subsection{\yy{SFRs in The Three Protoclusters}}
\yy{In Figure \ref{SFR}, our stacked samples shown in orange dots are compared with the stacked field field galaxies from
\citet{Sanders2021} shown in blue squares. The star-forming main sequence (SFMS) at $z=2.24$ derived by \citet{2014ApJS..214...15S} is also included here, and represented by the green shaded region with a 0.2 dex scatter.
We find that our stacked samples are generally consistant with the SFMS, 
although there is a slight elevated SFR in the lowest mass bin 5 (log$\frac{M_{\ast}}{M_{\odot}} \in [9.30, 9.75)$), which 
could be due to the detection limit of our grism data.
As the 3$\sigma$ SFR detection limit we plot in Figure \ref{SFR} as the gray dashed line, 
it suggests that the elevated SFR of the mass bin 5 could be an artifact of the limited sensitivity of our observations rather than a true enhancement in star formation activity. To be conservative we flag the lowest-mass bin as possibly elevated but not robustly above the SFMS.
The result of the SFRs closely following the SFMS is consistant with the results of Spiderweb protocluster \citep{2023MNRAS.518.1707P} and the NICE-COSMOS protoclusters \citep{2025A&A...694A.165T}.

}
\subsection{MZR in The Overdense Environments}

\begin{figure*}[!htb]
    \centering
    \includegraphics[width=1\textwidth]{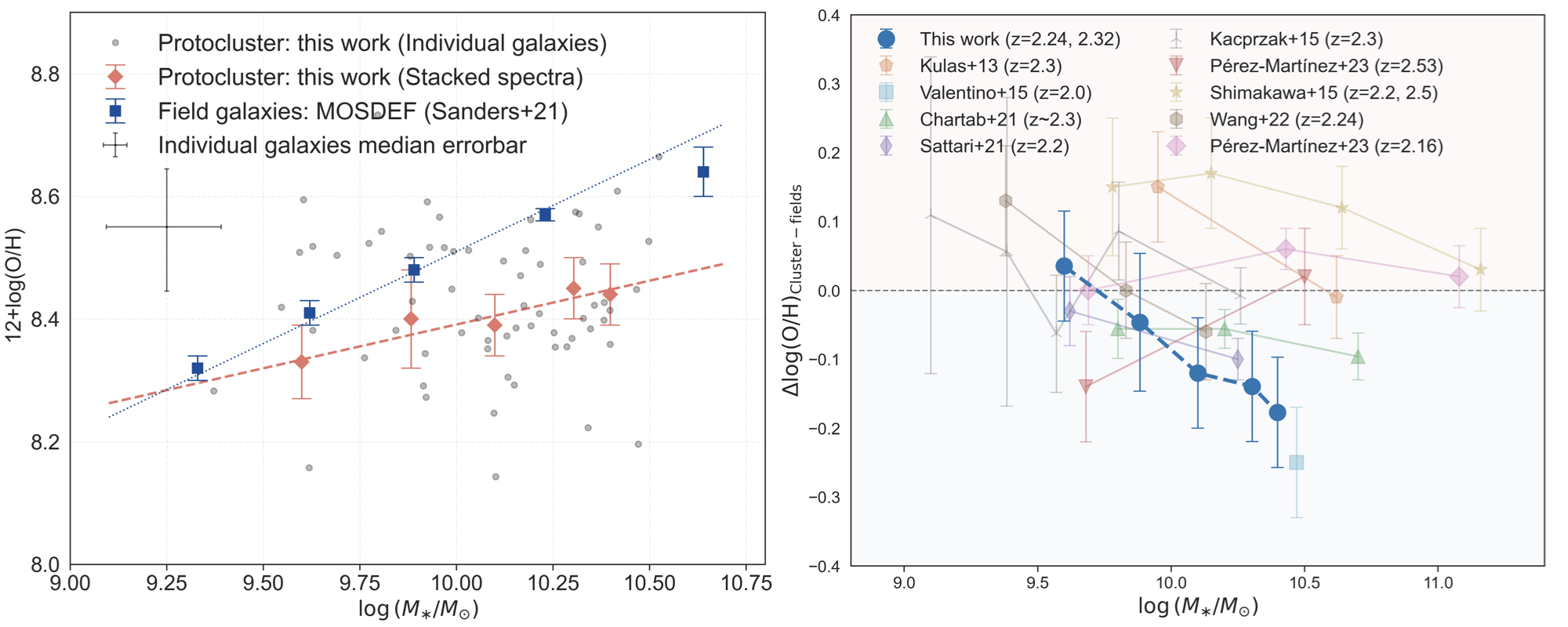}
    \caption{
    \textit{Left}: The MZR of protocluster member galaxies. Individual galaxies are shown in gray with median error bars in the upper-left corner. The stacked median values with 1$\sigma$ uncertainties are shown in red. The red dashed line represents the linear fit to the stacked points.
    \textit{Right}: Metallicity offset $\Delta {\rm log (O/H)} $ between protocluster galaxies and field galaxies as a function of stellar mass. The offset is calculated relative to the FMR from \cite{Sanders2021}.
    Our results are shown as dark red points connected by dashed lines. For comparison, we also include literature results at similar redshifts.
    }
    \label{fig:Mzr}
\end{figure*}

Our main results are presented in Figure \ref{fig:Mzr}, which shows the  stellar mass-metallicity relation relation and the metallicity offset compared to field galaxies as a function of stellar mass for protoclusters in the most overdense environments at cosmic noon ($z \sim 2.3$), when star formation activity in massive galaxies was at its peak.

In the left panel of Figure \ref{fig:Mzr}, the individual galaxies are shown in gray dots, with their median error bar plotted at the upper left of the diagram. 
The stacked results are shown in red, along with 1$\sigma$ uncertainties. 
The linear regression of the stacked data is displayed  by the red dotted line.
The derived MZR in the $M_{\ast}$ range of $[10^{9.0}, 10^{10.5}]$ is $12 + \log(\text{O/H}) = (6.96 \pm 0.13) + (0.143 \pm 0.017) \times \log(M_{\ast}/M_{\odot})$. The slope of our result is consistent with the previous work \citep{Wang_2022}.
\yy{Here we need to stress that different studies of the MZR are subject to several systematic effects that can mimic environmental signatures.  
Strong-line calibrations (e.g. N2, O3N2, R23) show systematic offsets and different sensitivities to 
ionization and ISM conditions, while selection criteria like emission-line S/N, UV/narrow-band selection, 
or photometric cuts will bias samples in SFR, dust and line strength, altering both MZR normalization and slope.  The choice of reference field sample and any SFR/FMR treatment also affects apparent offsets.  
To reduce these issues we compare our protocluster measurements to the MOSDEF field sample of \citet{Sanders2021} using the same B18 empirical calibration; nevertheless, residual systematic uncertainties remain, 
so reported metallicity differences should be interpreted as relative rather than absolute.}
We clearly observe that the slope of our result is significantly shallower than that of field galaxies, implying that the galaxies in the overdense environment at cosmic noon may 
be undergoing different evolution processes.

The fundamental metallicity relation (FMR) \citep{2010MNRAS.408.2115M,2013MNRAS.433.1425B, 2018ApJ...858...99S,curtiMassMetallicityFundamentalMetallicity2020} shows that at a fixed stellar mass, 
galaxies with higher SFRs have lower metallicities both in local and high-redshift universe (up to $ z \sim 2.5$). Recent JWST observations suggest that this relation may persist at even higher redshifts ($z \gtrsim 4$) \citep{Nakajima_23,Sarkar_25}.
Considering SFR differences between our samples and the field galaxies, we calculate the metallicity offset by using the equation from \cite{Sanders2021} as reference metallicity:
\begin{equation}\label{eq:mu}
\mu_{\alpha} \equiv \log(M_*/M_{\odot}) - \alpha\times\text{log}\left(\frac{\text{SFR}}{M_{\odot}~\text{yr}^{-1}}\right) .
\end{equation}
\begin{equation}\label{eq:fmr}
    12+\log(\mbox{O/H})= 8.80 + 0.188 y - 0.220 y^2 - 0.0531 y^3
\end{equation}
 Where $y = \mu_{0.60} - 10$. We then compare the offset between our samples and the field galaxies, represented by dark blue markers with a dashed line of the same color connecting them in the right panel of Figure \ref{fig:Mzr}.

 Our galaxies in the mass bin 5 (log$\frac{M_{\ast}}{M_{\odot}} \in [9.30, 9.75)$), galaxies show slight metal enrichment of $\Delta Z = 0.01 \pm 0.08$ dex, comparable to field galaxies. The mass bin 4 (log$\frac{M_{\ast}}{M_{\odot}} \in [9.75, 9.95)$) exhibits a small metal deficiency of $\Delta Z = -0.05 \pm 0.10$ dex, which becomes more pronounced in the mass bin 3 (log$\frac{M_{\ast}}{M_{\odot}} \in [9.95, 10.20)$: $\Delta Z = -0.12 \pm 0.08$ dex). The deficiency increases with mass, reaching $\Delta Z = -0.14 \pm 0.08$ dex in the mass bin 2 (log$\frac{M_{\ast}}{M_{\odot}} \in [10.20, 10.35]$) and $\Delta Z = -0.18 \pm 0.08$ dex in the mass bin 1 (log$\frac{M_{\ast}}{M_{\odot}} \in [10.35, 10.95]$).


We also include results from the literature \citep{kulasMASSMETALLICITYRELATION2013,2015ApJ...802L..26K, Shimakawa2015, 2015ApJ...801..132V, Sattari2021,chartabMOSDEFSurveyEnvironmental2021,Wang_2022, 2023MNRAS.518.1707P, perez-martinezEnhancedStarFormation2024}
that measured the offsets between field galaxies and galaxies in protoclusters at similar redshifts ($z \sim 2$) for comparison, as the right panel of Figure \ref{fig:Mzr}.
\yy{What should be kept in mind is that our metallicity are rely on [OIII]-dependent diagnostics, which is more sensitive to low-metallicity regimes.
In contrast, most literature results at similar redshifts \citet{kulasMASSMETALLICITYRELATION2013,2015ApJ...802L..26K, 2015ApJ...801..132V, Sattari2021,chartabMOSDEFSurveyEnvironmental2021, 2023MNRAS.518.1707P, perez-martinezEnhancedStarFormation2024} are based on [NII]/H$\alpha$ diagnostics, which are more sensitive to high-metallicity regimes.
The different choice of metallicity can result in systematic offsets, especially when comparing metallicities in the high mass bins.
Our samples could be more metal-poor compared to these studies.}
We observe that the metallicity offset in more massive galaxies, compared to field galaxies, is smaller than that in less massive galaxies, with the exception of the results from \cite{2015ApJ...801..132V} and \cite{perez-martinezEnhancedStarFormation2024}.
\cite{2015ApJ...801..132V} analyzed 6 galaxies in an X-ray-selected cluster, CL J1449+0856, at $z = 2$ and detected a metallicity deficiency of $\Delta Z = -0.25$ dex. Due to their small sample size, they did not detect a mass dependence on the metallicity offset. \cite{perez-martinezEnhancedStarFormation2024} analyzed 27 HAEs in the USS1558 protocluster at $z = 2.53$ and found metallicity deficit in low-mass galaxies, with no significant difference in high-mass galaxies. This may be because they did not correct the metallicity of field galaxies using the FMR, as their low-mass samples have significantly higher SFRs than the main sequence of galaxies shown in their Figure 5.

\cite{Shimakawa2015} compared their narrow-band selected galaxies with UV-selected field galaxies from \cite{2006ApJ...644..813E}, which could explain the metal enhancement in their samples, as UV-selected galaxies tend to be metal-poor due to selection bias. The results of \cite{kulasMASSMETALLICITYRELATION2013} may be influenced by the mass offset between their sample and field galaxies (see their Figure 2).

The results of \cite{Sattari2021} and \cite{chartabMOSDEFSurveyEnvironmental2021} are consistent with our findings, showing a similar mass dependence of the metallicity offset, with the metallicity deficit being more significant in high-mass galaxies than in low-mass galaxies.

\yy{We also note that some authors report no clear environmental shift in the MZR.  In particular, \citet{2015ApJ...802L..26K} found essentially identical cluster and field MZRs at $(z\sim2)$ using MOSFIRE/ZFIRE spectroscopy, while \citet{2023MNRAS.518.1707P} report only mild or mixed metallicity differences in the Spiderweb protocluster. 
These apparently discordant findings likely reflect differing sample definitions and target regions, metallicity calibrations, and the presence or absence of heavily obscured starbursts in the surveyed areas.  Such selection and environmental-definition differences can naturally reconcile null-results with the mass-dependent dilution/enrichment pattern we observe. 
}

\begin{figure}[htb]
    \centering
    \includegraphics[width=1\linewidth]{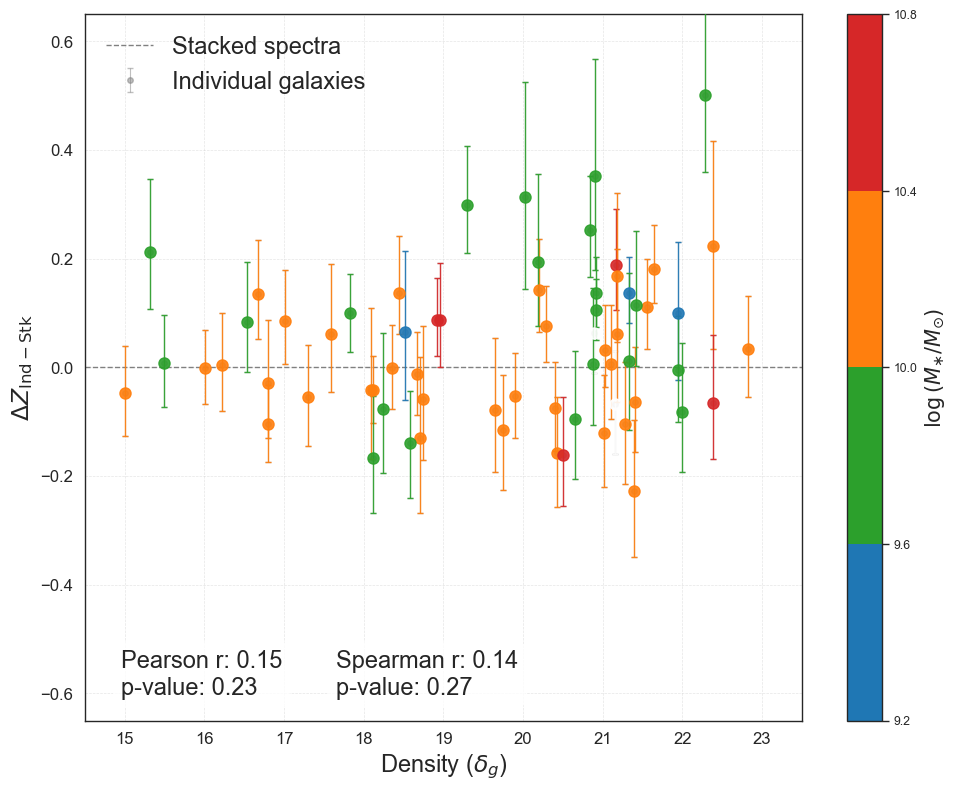}
    \caption{
    The metallicity offset of our sample galaxies relative to the derived MZR as a function of galaxy overdensity $\delta_g$, which
    is calculated using $\delta_g = \Sigma_{\rm group}/\Sigma_{\rm field} -1$. We performed Pearson and Spearman rank correlation tests to examine the relationship between galaxy overdensity and metallicity offset. 
    Both tests indicate no significant correlation between galaxy overdensity and the metallicity offset. 
    }
    \label{density}
\end{figure}

To further examine whether galaxy overdensity affects metallicity, we computed the metallicity offset relative to the derived MZR as a function of the galaxy overdensity ($\delta_g$), as shown in Figure \ref{density}.
The galaxy overdensity, $\delta_g$, is estimated from the surface density of H$\alpha$ emitters (HAEs) in BOSS1244 and BOSS1542, or Ly$\alpha$ emitters (LAEs) in BOSS1441, using the relation $\delta_g = \Sigma_{\rm group}/\Sigma_{\rm field} - 1$. Here, $\Sigma_{\rm_group}$ represents the surface density of HAEs or LAEs per arcmin$^2$ within the protoclusters, while $\Sigma_{\rm_field}$ refers to the corresponding surface density in random fields at similar redshifts.
\yy{The luminosity limits for selecting HAEs and LAEs in the protoclusters are $L_{\rm H\alpha} = 1.00 \times 10^{42}$ erg s$^{-1}$ and $L_{\rm Ly\alpha} = 1.56 \times 10^{42}$ erg s$^{-1}$, respectively \citep{MAMMOTH2017B,2021ApJ...915...32S}.
}
We adopt a field surface density of $\Sigma_{\rm field} = 7.25 \times 10^{-2}$ HAEs per arcmin$^2$, following \citet{2021ApJ...915...32S}.
The field surface density of LAEs is estimated to be $\Sigma_{\rm field} = 1.67 \times 10^{-2}$ LAEs per arcmin$^2$, following \citet{MAMMOTH2017B}.
We then generate the $\delta_g$ maps of the three protoclusters by applying the 2D kernel density estimation (KDE) to the HAE or LAE maps, using a Gaussian kernel of 150″ ($\rm \sim 4.0 cMpc$) smoothing scale.  
To assess the relationship between galaxy overdensity and metallicity offset, we performed both Pearson and Spearman rank correlation tests. The resulting correlation coefficients are r = 0.15 (p = 0.25) and r = 0.13 (p = 0.29), respectively.
These results indicate that there is no statistically significant correlation between galaxy overdensity and metallicity offset in our sample, \yy{which is different with the findings of \citet{2023MNRAS.518.1707P}.
They found a mild anti-correlation between the metallicity offset and local density in PKS1138 at $z=2.16$. However, in \citet{perez-martinezEnhancedStarFormation2024}, their results show no significant trends of metallicity offset on local density in USS1558 at $z=2.53$.
} This may be attributed to the fact that the galaxies in our sample are primarily located in the dense core regions of protoclusters, where variations in local density are relatively small. Additionally, the apparent scatter across different densities is likely a consequence of the limited sample size.

\section{Discussion}\label{discuss}
We have presented one of the first MZR measurements in multiple massive protoclusters at cosmic noon. By synthesizing our results with those from the literature into a mass-metallicity offset diagram, we identify a trend: massive protocluster galaxies tend to be more metal-poor than their field counterparts, while low-mass galaxies exhibit comparable or slightly enhanced metallicities. 
These results may be explained by the interplay of two physical processes: (1) the presence of small-scale ($\sim$kiloparsec) filamentary gas streams, and (2) the efficient recycling of feedback-driven outflows. 
\begin{enumerate}
    \item  \textbf{Small-scale gas stream:}  
    \citet{Wangk_23} used EAGLE simulation to investigate the environmental impact on MZR at $z \lesssim 2$. They found that central galaxies in protoclusters at $z \sim 2$ typically reside in more massive halos, experience enhanced cold-mode gas accretion, and consequently show lower metallicities, consistent with our observations.
    \citet{galarraga-espinosaFlowsGalaxiesDependence2023b} used the output of the TNG50-1 simulation to examine the relationship between galaxy connectivity 
    (i.e., the number of small-scale filamentary streams connected to a galaxy) and cosmic environment at $z = 2$. 
    They also explored the impact of connectivity on the SFR. Their findings 
    indicate that galaxies in cosmic web knots (i.e., protoclusters or clusters) have higher connectivity than field galaxies, 
    meaning that galaxies in these environments at $z = 2$ can accrete more fresh gas from the cosmic web, 
    leading to the dilution of their metallicity. Interestingly, the galaxy connectivity is found to significantly enhance SFR in low-mass galaxies ($M_{\ast} < 10^{9.5} M_{\odot}/h$, or $M_{\ast} < 10^{9.65} M_{\odot}$ when h = 0.7), but has no significant effect on high-mass galaxies,
     which is consistent with our results (Figure \ref{SFR}). This scenario suggests a transition in dominant accretion
      modes from cold-mode to hot-mode accretion in galaxies of different masses using the $M_{\ast} - M_{Halo}$ relation 
      from \cite{mosterCONSTRAINTSRELATIONSHIPStelLAR2010} with a dark matter halo threshold mass $M_{shock} \approx 10^{12} M_{\odot}$ from \cite{dekelGalaxyBimodalityDue2006}. 
      While the gas streams can account for the observed metallicity deficit and elevated SFR in our results, another process is needed to explain the lack of significant metallicity deficit in low-mass galaxies.
    \item \textbf{Efficient recycling of feedback-driven winds:} 
    \citet{Ma2016} used simulations and found galaxies above $M_*=10^6M_\odot$ are able to retain a large fraction of their metals as metal-enriched winds are unable to fully escape and are instead recycled back into the galaxy. \citet{zheng2023} detected emission lines from ionized carbon that extended 100 kpc from a massive galaxy at $z=2.3$, with kinematics consistent with an inspiraling stream, indicating the presence of chemically enriched gas in the circumgalactic medium.The combination of inspiraling motion and high carbon abundance suggests that the gas was previously ejected and is now reaccreting, providing direct evidence for gas recycling.
    \cite{oppenheimerMassMetalEnergy2008} used cosmological hydrodynamic simulations and found that the recycling of gas and metals ejected by momentum-driven winds is primarily influenced by environmental density. In denser environments, the recycling of wind material is more efficient, which facilitates the enrichment of galaxy metallicity. This recycling effect is more pronounced in low-mass galaxies, as they and \citet{el-badryGasKinematicsMorphology2018} found high-mass galaxies are inherently less affected by ejective feedback mechanisms due to their deeper potential wells.
    \citet{Alcazar_17} also showed that in lower-mass galaxies, a greater proportion of the gas supply comes from wind recycling, with these systems re-accreting more gas relative to their stellar mass.

\end{enumerate}
\yy{Finally, we note two observational subtleties that help reconcile the apparent tension with simple FMR expectations and with some literature results. 
 First, our dataset is emission-line selected and comprises galaxies with only moderate dust extinction; it therefore lacks the heavily obscured starburst population that some protocluster studies target.  
 If such dusty starbursts dominate certain core regions, they could produce elevated SFRs at the high-mass end and thereby alter the comparison.  Second, inflow and recycling efficiencies vary systematically with 
 halo mass and local density: pristine cold streams preferentially feed massive galaxies located at filament nodes and can dilute nebular abundances without necessarily triggering 
 an immediate SFR boost (e.g., when the accreted gas has high angular momentum or is temporarily heated).  By contrast, low-mass satellites inhabit very dense sub-environments 
 where outflows are more readily trapped and recycled, promoting both higher retained metallicity and short-term SFR.  The same deep potentials that enhance recycling can also impede some cold inflows (or strip them), so inflow and recycling efficiencies 
 are anti-correlated with halo mass and local density.  These competing, mass-dependent effects naturally produce dilution at the massive end (with little SFR response) and enhanced metallicity at the low-mass end, and are consistent with a 
 crossover mass near $M_{\rm crossover}\sim10^{9.8}\,M_\odot$.  We stress, however, that confirming this picture requires complementary data (dust-sensitive SFR tracers and direct gas-mass measures) which are beyond the present dataset.
 Last but not least, we note a straightforward observational route to test the cold-stream + recycling picture.  Keck/KCWI can reveal extended Ly$\alpha$ nebulae and spatially resolved interstellar absorption, which trace cool CGM gas, inflow kinematics and metal content.  These diagnostics can help separate pristine, low-metallicity inflows from metal-rich recycled gas via morphology, velocity structure and absorption strength.  By contrast, obtaining resolved H$\alpha$ maps for our targets with HST grisms is not feasible at these redshifts.  Thus, combining KCWI Ly$\alpha$ mapping with dust-sensitive SFR tracers and direct cold-gas measurements (e.g.\ ALMA/NOEMA) offers the most practical near-term path to distinguish the roles of cold accretion and wind recycling.
}




\section{Conclusion}\label{con}

In this paper, we studied the MZR of galaxies in the overdense environments at cosmic noon, using a sample of 63 galaxies from the protoclusters BOSS1244, BOSS1441, and BOSS1542.
Using grism data acquired by MAMMOTH-Grism, and our forward-modeling Bayesian inference method, we obtained the gas-phase metallicities and SFRs of these galaxies.
By taking advantage of existing multi-wavelength imaging data, 
 we estimated the stellar masses of the sample galaxies. Using the MEx diagram, we excluded AGN contamination from the galaxy sample, ensuring that primarily ionized by massive stars in \HII regions rather than AGNs.
We divided our sample into five mass bins and stacked the grism spectra in each bin to derive the MZR of galaxies in overdense environments at cosmic noon.

Our main results are as follows:
\begin{enumerate}
    \item The MZR of galaxies in the most overdense environment at cosmic noon is shallower than that of field galaxies, with a slope of $0.143 \pm 0.017$.
    \item The metallicity offset between galaxies in the protoclusters and field galaxies is mass-dependent, with more massive galaxies showing a more significant metallicity deficit than less massive galaxies.
    \item The metallicity deficit in high mass galaxies is more significant than in low mass galaxies, which may be caused by the combination of small-scale gas streams and efficient recycling of feedback-driven winds.
    \item The density of the local environment does not significantly affect the metallicity of galaxies in our sample.
\end{enumerate}

We thank the anonymous referee for a careful read and useful comments that helped improve the clarity of this paper.
We thank Hang Zhou, Xunda Sun, and Hanwen Sun for helpful discussions. 
\yy{Some of the data presented in this article were obtained from the Mikulski Archive for Space Telescopes (MAST) at the Space Telescope Science Institute. The specific observations analyzed can be accessed via \dataset[doi:10.17909/4012-2q76]{[https://doi.org/10.17909/4012-2q76]}.}
This work is supported by the China Manned Space Program with grant no. CMS-CSST-2025-A06, the National Natural Science Foundation of China (grant 12373009), the CAS Project for Young Scientists in Basic Research Grant No. YSBR-062, the Fundamental Research Funds for the Central Universities, the Xiaomi Young Talents Program.  CWT is supported by NSFC 12588202. This work is also supported by NASA through HST grant HST-GO-16276 and HST-GO-17159. D.D.S. acknowledges support from the National Science Foundation of China (Grant No. 12303015), the National Science Foundation of Jiangsu Province (Grant No. BK20231106). X.Z.Z. is supported by the National Key Research and Development Program of China (2023YFA1608100), the National Science Foundation of China (NSFC, Grant No. 12233005), the China Manned Space Program with grants nos. CMS-CSST-2025-A08 and CMS-CSST-2025-A20, and  Office of Science and Technology, Shanghai Municipal Government (grant Nos. 24DX1400100，ZJ2023-ZD-001).

%
\vspace{5mm}
\facilities{HST(WFC3), LBT(LBC). CFHT(WIRCam)}


\software{Source Extractor \citep{1996A&AS..117..393B},
            Astropy \citep{2013ascl.soft04002G},  
            Grizli \citep{2019ascl.soft05001B},
            T-PHOT \citep{Merlin.2016},
            BAGPIPES \citep{2018MNRAS.480.4379C},
            emcee \citep{emcee},
            LMFIT \citep{lmfit}
            }


\newpage
\appendix
\section{Individual Galaxies Properties}
{
\tabletypesize{\scriptsize}
\tabcolsep=4pt
\startlongtable
\begin{deluxetable}{ccccccccccccc}   \tablecolumns{13}
\tablewidth{0pt}
\tablecaption{Measured Properties of Individual Sources in the Protoclusters Member Galaxy Sample}
\tablehead{
    \colhead{ID} & 
    \colhead{R.A.} & 
    \colhead{Dec.} & 
    \colhead{$z_\mathrm{grism}$} & 
    \multicolumn{5}{c}{Observed emission line fluxes [$10^{-17}$\Funit]} &
    \multicolumn{4}{c}{Derived physical properties} \\
    & [deg.] & [deg.] &  & \colhead{$f_{\rm [OII]}$} & 
    \colhead{$f_{\rm [NeIII]}$} & 
    \colhead{$f_{\rm H\gamma}$} & 
    \colhead{$f_{\rm H\beta}$} & 
    \colhead{$f_{\rm [OIII]}$} & 
    \colhead{$\log(M_{\ast}/M_{\odot})$} & 
    \colhead{A$_{\rm V}$} & 
    \colhead{SFR [\Msun/yr]} &
    \colhead{$12+\log({\rm O/H})$} 
    }
\startdata
\multicolumn{13}{c}{Mass bin 5}   \\
\noalign{\smallskip}\hline\noalign{\smallskip}
1394 & 190.877806 & 35.898194 & 2.213390 & $5.19 \pm 0.61$ & $0.45 \pm 0.86$ & $1.07 \pm 0.53$ & $2.58 \pm 0.34$ & $9.15 \pm 0.41$ & $9.59_{-0.12}^{+0.13}$ & $0.49_{-0.33}^{+0.46}$ & $22.43_{-7.41}^{+16.21}$ & $8.51_{-0.07}^{+0.06}$ \\
2731 & 190.902840 & 35.937397 & 2.286767 & $27.88 \pm 4.95$ & $7.94 \pm 7.74$ & {...} & $1.31 \pm 3.22$ & $22.58 \pm 3.65$ & $9.63_{-0.21}^{+0.16}$ & $0.26_{-0.19}^{+0.42}$ & $54.55_{-15.89}^{+33.08}$ & $8.52_{-0.10}^{+0.09}$ \\
3032 & 220.359566 & 40.016983 & 2.337503 & $3.96 \pm 1.10$ & $1.62 \pm 1.91$ & $0.77 \pm 1.19$ & $1.23 \pm 0.81$ & $7.85 \pm 1.35$ & $9.55_{-0.31}^{+0.24}$ & $0.80_{-0.58}^{+1.00}$ & $23.59_{-12.01}^{+50.06}$ & $8.42_{-0.15}^{+0.13}$ \\
3745 & 220.404855 & 40.038908 & 2.246653 & $5.52 \pm 1.52$ & ... & $1.94 \pm 1.38$ & $3.31 \pm 1.11$ & $21.68 \pm 1.53$ & $9.37_{-0.42}^{+0.22}$ & $1.19_{-0.81}^{+1.02}$ & $66.30_{-40.98}^{+153.64}$ & $8.28_{-0.13}^{+0.12}$ \\
3904 & 220.356893 & 40.013687 & 2.304711 & $14.57 \pm 1.90$ & $6.29 \pm 2.77$ & $2.15 \pm 1.97$ & $4.13 \pm 1.44$ & $29.90 \pm 1.73$ & $9.63_{-0.21}^{+0.16}$ & $0.33_{-0.24}^{+0.43}$ & $48.08_{-12.78}^{+34.05}$ & $8.38_{-0.07}^{+0.07}$ \\
2808 & 235.703888 & 38.869335 & 2.312362 & $2.03 \pm 0.55$ & $0.74 \pm 0.79$ & $0.88 \pm 0.51$ & $0.54 \pm 0.38$ & $2.91 \pm 0.48$ & $9.69_{-0.11}^{+0.13}$ & $0.67_{-0.49}^{+0.98}$ & $9.58_{-4.48}^{+18.78}$ & $8.50_{-0.14}^{+0.11}$ \\
3443 & 235.703550 & 38.891068 & 2.284314 & $5.54 \pm 1.75$ & {...} & $0.78 \pm 1.41$ & {...} & $7.80 \pm 1.48$ & $9.60_{-0.21}^{+0.15}$ & $0.97_{-0.70}^{+1.15}$ & $46.45_{-29.77}^{+134.66}$ & $8.59_{-0.21}^{+0.17}$ \\
3534 & 235.692625 & 38.893567 & 2.204170 & $4.24 \pm 0.87$ & $2.48 \pm 1.04$ & $1.55 \pm 0.76$ & $2.62 \pm 0.62$ & $23.86 \pm 0.84$ & $9.62_{-0.06}^{+0.06}$ & $0.94_{-0.66}^{+0.95}$ & $42.23_{-22.09}^{+82.66}$ & $8.16_{-0.10}^{+0.10}$ \\
\noalign{\smallskip}\hline\noalign{\smallskip}
\multicolumn{13}{c}{Mass bin 4}   \\
\noalign{\smallskip}\hline\noalign{\smallskip}
1071 & 190.867795 & 35.884012 & 2.236753 & $3.62 \pm 0.96$ & ... & ... & $0.69 \pm 0.71$ & $7.22 \pm 0.83$ & $9.84_{-0.20}^{+0.13}$ & $0.75_{-0.55}^{+1.03}$ & $17.10_{-8.37}^{+37.44}$ & $8.38_{-0.14}^{+0.12}$ \\
1397 & 190.884983 & 35.898280 & 2.318674 & $4.37 \pm 1.09$ & $0.49 \pm 1.49$ & ... & $0.71 \pm 0.74$ & $9.50 \pm 0.98$ & $9.76_{-0.20}^{+0.16}$ & $0.69_{-0.50}^{+0.94}$ & $20.52_{-9.32}^{+37.95}$ & $8.34_{-0.13}^{+0.11}$ \\
1573 & 190.882538 & 35.905175 & 2.227217 & $2.52 \pm 0.76$ & $1.68 \pm 1.03$ & $1.86 \pm 0.78$ & ... & $6.02 \pm 0.91$ & $9.80_{-0.20}^{+0.15}$ & $1.59_{-0.95}^{+1.01}$ & $97.12_{-76.58}^{+273.30}$ & $8.73_{-0.26}^{+0.14}$ \\
1890 & 190.865939 & 35.913638 & 2.250783 & $6.71 \pm 1.74$ & $0.66 \pm 2.52$ & $0.11 \pm 1.71$ & $1.77 \pm 1.22$ & $19.71 \pm 1.59$ & $9.92_{-0.23}^{+0.20}$ & $0.86_{-0.61}^{+1.03}$ & $41.54_{-21.29}^{+92.25}$ & $8.27_{-0.12}^{+0.11}$ \\
2474 & 190.905843 & 35.929694 & 2.191480 & $5.00 \pm 1.55$ & ... & $1.80 \pm 1.54$ & $1.40 \pm 1.01$ & $8.97 \pm 1.20$ & $9.89_{-0.17}^{+0.17}$ & $0.87_{-0.63}^{+1.19}$ & $25.78_{-13.74}^{+69.28}$ & $8.43_{-0.16}^{+0.13}$ \\
3331 & 190.862922 & 35.967686 & 2.223014 & $4.12 \pm 0.99$ & ... & $0.13 \pm 0.90$ & $0.86 \pm 0.63$ & $10.52 \pm 0.78$ & $9.91_{-0.17}^{+0.16}$ & $0.73_{-0.53}^{+0.95}$ & $19.28_{-8.88}^{+37.05}$ & $8.29_{-0.11}^{+0.10}$ \\
3382 & 220.390044 & 40.026162 & 2.307205 & $10.29 \pm 1.21$ & ... & $1.71 \pm 1.23$ & $2.61 \pm 0.79$ & $10.29 \pm 0.97$ & $9.81_{-0.19}^{+0.19}$ & $0.16_{-0.12}^{+0.24}$ & $24.71_{-4.75}^{+8.67}$ & $8.54_{-0.07}^{+0.06}$ \\
3028 & 220.414100 & 39.994593 & 2.261937 & $8.99 \pm 2.31$ & $2.78 \pm 4.77$ & ... & $3.39 \pm 1.62$ & $9.73 \pm 1.84$ & $9.92_{-0.11}^{+0.14}$ & $0.58_{-0.42}^{+0.84}$ & $38.09_{-17.07}^{+57.02}$ & $8.59_{-0.13}^{+0.11}$ \\
4179 & 220.368649 & 40.018750 & 2.320003 & $18.98 \pm 4.22$ & ... & $4.84 \pm 4.65$ & $10.42 \pm 2.88$ & $36.24 \pm 3.22$ & $9.93_{-0.20}^{+0.21}$ & $0.78_{-0.53}^{+0.76}$ & $137.57_{-67.58}^{+201.82}$ & $8.52_{-0.11}^{+0.09}$ \\
4508 & 220.390078 & 40.026171 & 2.306926 & $11.61 \pm 1.11$ & $1.44 \pm 1.78$ & $3.04 \pm 1.15$ & $0.83 \pm 0.77$ & $9.26 \pm 0.95$ & $9.77_{-0.20}^{+0.19}$ & $0.09_{-0.07}^{+0.14}$ & $21.09_{-3.15}^{+4.47}$ & $8.52_{-0.06}^{+0.05}$ \\
4971 & 220.404805 & 40.038917 & 2.246954 & $8.36 \pm 1.67$ & $1.26 \pm 2.17$ & $1.12 \pm 1.47$ & $3.49 \pm 1.17$ & $22.45 \pm 1.72$ & $9.92_{-0.14}^{+0.09}$ & $0.74_{-0.51}^{+0.76}$ & $48.02_{-22.39}^{+71.96}$ & $8.34_{-0.10}^{+0.10}$ \\
1292 & 235.712784 & 38.831004 & 2.185424 & $6.34 \pm 1.71$ & ... & $3.74 \pm 1.78$ & $2.31 \pm 1.31$ & $10.97 \pm 1.49$ & $9.88_{-0.18}^{+0.17}$ & $0.72_{-0.51}^{+0.91}$ & $32.83_{-16.08}^{+60.26}$ & $8.50_{-0.14}^{+0.11}$ \\
\noalign{\smallskip}\hline\noalign{\smallskip}
\multicolumn{13}{c}{Mass bin 3}   \\
\noalign{\smallskip}\hline\noalign{\smallskip}
381 & 190.909141 & 35.854547 & 2.179024 & $5.38 \pm 0.95$ & ... & $0.33 \pm 0.90$ & $1.56 \pm 0.63$ & $6.88 \pm 0.76$ & $9.99_{-0.16}^{+0.18}$ & $0.35_{-0.26}^{+0.51}$ & $14.65_{-4.50}^{+12.06}$ & $8.51_{-0.09}^{+0.08}$ \\
613 & 190.921119 & 35.865813 & 2.243770 & $5.16 \pm 1.26$ & $0.38 \pm 2.23$ & $2.29 \pm 1.22$ & $2.05 \pm 0.82$ & $11.99 \pm 0.97$ & $10.06_{-0.19}^{+0.13}$ & $0.78_{-0.55}^{+0.85}$ & $30.88_{-15.21}^{+52.94}$ & $8.40_{-0.11}^{+0.10}$ \\
769 & 190.909740 & 35.871348 & 2.241579 & $15.72 \pm 2.34$ & $6.72 \pm 4.83$ & $1.17 \pm 2.80$ & $4.52 \pm 1.61$ & $33.02 \pm 2.46$ & $10.13_{-0.17}^{+0.15}$ & $0.38_{-0.28}^{+0.50}$ & $51.02_{-15.22}^{+42.25}$ & $8.37_{-0.08}^{+0.08}$ \\
996 & 190.873758 & 35.880726 & 2.320312 & $2.82 \pm 0.89$ & $2.01 \pm 1.09$ & $1.31 \pm 0.79$ & $1.47 \pm 0.69$ & $10.41 \pm 0.97$ & $10.15_{-0.20}^{+0.17}$ & $1.26_{-0.86}^{+1.16}$ & $38.08_{-24.51}^{+106.27}$ & $8.29_{-0.15}^{+0.14}$ \\
1335 & 190.871757 & 35.895697 & 2.235818 & $10.42 \pm 1.64$ & $3.06 \pm 3.35$ & $1.65 \pm 1.66$ & $2.70 \pm 0.79$ & $9.28 \pm 0.93$ & $10.19_{-0.16}^{+0.12}$ & $0.21_{-0.16}^{+0.34}$ & $22.79_{-5.32}^{+10.81}$ & $8.56_{-0.07}^{+0.07}$ \\
1435 & 190.869813 & 35.900036 & 2.209610 & $5.09 \pm 0.70$ & $1.32 \pm 0.84$ & $1.22 \pm 0.61$ & $2.72 \pm 0.46$ & $15.03 \pm 0.61$ & $10.08_{-0.12}^{+0.10}$ & $0.79_{-0.52}^{+0.65}$ & $33.16_{-15.52}^{+39.79}$ & $8.35_{-0.09}^{+0.08}$ \\
1464 & 190.873403 & 35.901270 & 2.211873 & $4.40 \pm 0.67$ & $2.05 \pm 0.88$ & ... & $2.32 \pm 0.44$ & $8.26 \pm 0.53$ & $10.18_{-0.17}^{+0.13}$ & $0.65_{-0.43}^{+0.59}$ & $23.90_{-9.94}^{+24.48}$ & $8.51_{-0.08}^{+0.07}$ \\
1467 & 190.869254 & 35.901455 & 2.210074 & $2.94 \pm 0.51$ & $2.33 \pm 0.66$ & $1.02 \pm 0.45$ & $2.12 \pm 0.33$ & $14.16 \pm 0.44$ & $10.10_{-0.19}^{+0.14}$ & $1.23_{-0.77}^{+0.87}$ & $40.97_{-24.29}^{+72.34}$ & $8.25_{-0.10}^{+0.10}$ \\
2327 & 190.927963 & 35.926164 & 2.238317 & $13.91 \pm 2.04$ & ... & $0.31 \pm 1.94$ & $5.18 \pm 1.41$ & $29.94 \pm 1.82$ & $10.15_{-0.16}^{+0.16}$ & $0.46_{-0.33}^{+0.54}$ & $52.12_{-17.67}^{+48.55}$ & $8.39_{-0.08}^{+0.08}$ \\
3385 & 190.853893 & 35.971115 & 2.300365 & $4.75 \pm 0.87$ & ... & $1.29 \pm 0.88$ & $0.79 \pm 0.68$ & $6.85 \pm 0.87$ & $10.17_{-0.17}^{+0.16}$ & $0.36_{-0.27}^{+0.54}$ & $14.88_{-4.74}^{+13.49}$ & $8.47_{-0.10}^{+0.09}$ \\
3496 & 190.846119 & 35.977685 & 2.235796 & $7.51 \pm 1.34$ & ... & $0.78 \pm 1.28$ & $0.71 \pm 0.83$ & $6.60 \pm 1.08$ & $10.03_{-0.16}^{+0.16}$ & $0.27_{-0.20}^{+0.45}$ & $14.89_{-4.32}^{+9.83}$ & $8.51_{-0.10}^{+0.08}$ \\
3516 & 190.851775 & 35.979114 & 2.233867 & $5.49 \pm 1.23$ & $1.34 \pm 1.79$ & $1.01 \pm 1.15$ & $3.17 \pm 0.80$ & $13.18 \pm 1.06$ & $9.99_{-0.15}^{+0.15}$ & $1.00_{-0.65}^{+0.86}$ & $47.55_{-26.50}^{+82.82}$ & $8.45_{-0.11}^{+0.09}$ \\
4479 & 220.422619 & 40.054613 & 2.340896 & $5.58 \pm 1.67$ & $2.27 \pm 2.01$ & $1.25 \pm 1.45$ & $4.93 \pm 1.19$ & $22.52 \pm 2.06$ & $10.01_{-0.15}^{+0.11}$ & $1.76_{-1.03}^{+1.01}$ & $170.73_{-123.28}^{+381.70}$ & $8.38_{-0.15}^{+0.12}$ \\
4582 & 220.423681 & 40.057109 & 2.276418 & $3.93 \pm 0.85$ & $2.42 \pm 1.46$ & $0.76 \pm 0.87$ & $2.10 \pm 0.60$ & $13.64 \pm 0.74$ & $10.13_{-0.36}^{+0.25}$ & $1.00_{-0.67}^{+0.91}$ & $36.32_{-20.16}^{+71.14}$ & $8.31_{-0.11}^{+0.10}$ \\
4640 & 220.431501 & 40.058304 & 2.310424 & $1.68 \pm 0.49$ & $0.17 \pm 0.57$ & $0.24 \pm 0.44$ & $1.73 \pm 0.49$ & $5.26 \pm 0.80$ & $9.97_{-0.12}^{+0.11}$ & $1.93_{-1.01}^{+1.04}$ & $64.53_{-47.41}^{+154.73}$ & $8.52_{-0.16}^{+0.12}$ \\
3750 & 220.426202 & 40.010749 & 2.271204 & $8.42 \pm 2.33$ & $2.52 \pm 4.19$ & $1.87 \pm 2.77$ & $4.26 \pm 1.41$ & $19.69 \pm 1.82$ & $10.17_{-0.16}^{+0.16}$ & $0.98_{-0.66}^{+0.96}$ & $67.72_{-37.67}^{+136.89}$ & $8.42_{-0.13}^{+0.11}$ \\
4189 & 220.397710 & 40.018591 & 2.287011 & $6.64 \pm 1.30$ & $5.44 \pm 1.87$ & $1.71 \pm 1.37$ & $5.61 \pm 1.15$ & $45.30 \pm 1.45$ & $10.10_{-0.16}^{+0.14}$ & $1.35_{-0.88}^{+0.97}$ & $135.82_{-85.63}^{+277.91}$ & $8.14_{-0.10}^{+0.11}$ \\
1203 & 235.711177 & 38.828104 & 2.237434 & $21.04 \pm 6.62$ & $10.95 \pm 6.78$ & $1.12 \pm 5.57$ & ... & $29.22 \pm 6.21$ & $9.96_{-0.14}^{+0.16}$ & $0.94_{-0.68}^{+1.14}$ & $147.57_{-91.60}^{+417.61}$ & $8.57_{-0.22}^{+0.17}$ \\
1704 & 235.714941 & 38.841340 & 2.240712 & $5.80 \pm 1.39$ & ... & $2.35 \pm 1.41$ & ... & $16.63 \pm 1.24$ & $10.12_{-0.15}^{+0.17}$ & $1.53_{-0.96}^{+0.92}$ & $127.53_{-96.45}^{+368.28}$ & $8.49_{-0.19}^{+0.19}$ \\
4943 & 235.684359 & 38.932112 & 2.236404 & $8.24 \pm 1.02$ & $0.75 \pm 1.41$ & $4.94 \pm 0.93$ & $6.53 \pm 0.71$ & $32.73 \pm 0.91$ & $10.19_{-0.07}^{+0.19}$ & $1.47_{-0.68}^{+0.59}$ & $167.76_{-93.41}^{+168.72}$ & $8.39_{-0.10}^{+0.08}$ \\
6235 & 235.667089 & 39.018249 & 2.245371 & $10.13 \pm 1.46$ & $0.16 \pm 1.89$ & ... & $2.91 \pm 1.13$ & $22.57 \pm 1.25$ & $10.08_{-0.07}^{+0.08}$ & $0.41_{-0.30}^{+0.51}$ & $35.60_{-11.14}^{+31.13}$ & $8.37_{-0.08}^{+0.08}$ \\
\noalign{\smallskip}\hline\noalign{\smallskip}
\multicolumn{13}{c}{Mass bin 2}   \\
\noalign{\smallskip}\hline\noalign{\smallskip}
313 & 190.902894 & 35.851659 & 2.227321 & $8.56 \pm 1.25$ & ... & $1.29 \pm 1.31$ & $4.50 \pm 0.78$ & $24.56 \pm 0.93$ & $10.26_{-0.13}^{+0.11}$ & $0.77_{-0.51}^{+0.66}$ & $54.48_{-25.05}^{+66.26}$ & $8.35_{-0.09}^{+0.08}$ \\
1112 & 190.879330 & 35.886279 & 2.300383 & $8.89 \pm 1.53$ & ... & ... & $4.46 \pm 1.16$ & $13.09 \pm 1.42$ & $10.31_{-0.10}^{+0.11}$ & $0.51_{-0.36}^{+0.56}$ & $44.83_{-17.05}^{+42.71}$ & $8.57_{-0.09}^{+0.08}$ \\
1998 & 190.868202 & 35.916375 & 2.213119 & $10.84 \pm 2.36$ & ... & $2.20 \pm 2.70$ & $5.58 \pm 1.62$ & $30.41 \pm 1.93$ & $10.30_{-0.10}^{+0.10}$ & $0.93_{-0.63}^{+0.81}$ & $81.86_{-44.10}^{+135.17}$ & $8.37_{-0.11}^{+0.10}$ \\
2588 & 190.872857 & 35.933017 & 2.244739 & $9.85 \pm 1.39$ & $4.32 \pm 2.40$ & $1.84 \pm 1.44$ & $3.96 \pm 1.32$ & $12.46 \pm 1.77$ & $10.32_{-0.11}^{+0.10}$ & $0.33_{-0.24}^{+0.42}$ & $34.47_{-10.30}^{+23.95}$ & $8.57_{-0.09}^{+0.08}$ \\
2992 & 190.843993 & 35.949979 & 2.347807 & $6.47 \pm 0.93$ & $0.65 \pm 1.35$ & $1.55 \pm 0.90$ & $3.71 \pm 0.74$ & $13.97 \pm 1.49$ & $10.22_{-0.16}^{+0.14}$ & $0.79_{-0.49}^{+0.60}$ & $50.39_{-23.51}^{+55.30}$ & $8.49_{-0.09}^{+0.08}$ \\
3061 & 190.838668 & 35.953087 & 2.225519 & $12.00 \pm 1.37$ & ... & $0.69 \pm 1.27$ & $5.33 \pm 0.90$ & $25.51 \pm 1.06$ & $10.21_{-0.17}^{+0.11}$ & $0.47_{-0.33}^{+0.48}$ & $46.99_{-15.70}^{+38.02}$ & $8.41_{-0.07}^{+0.07}$ \\
3495 & 190.853920 & 35.977367 & 2.231189 & $7.72 \pm 1.13$ & ... & $1.99 \pm 1.04$ & $2.26 \pm 0.63$ & $14.28 \pm 0.77$ & $10.33_{-0.16}^{+0.15}$ & $0.31_{-0.23}^{+0.43}$ & $22.02_{-5.58}^{+14.60}$ & $8.40_{-0.07}^{+0.07}$ \\
3652 & 220.411198 & 40.035707 & 2.180092 & $2.32 \pm 0.66$ & $0.17 \pm 0.80$ & $0.29 \pm 0.59$ & $1.57 \pm 0.46$ & $11.34 \pm 0.59$ & $10.34_{-0.07}^{+0.12}$ & $1.48_{-0.94}^{+1.13}$ & $39.44_{-26.32}^{+103.73}$ & $8.22_{-0.13}^{+0.12}$ \\
1291 & 235.712874 & 38.830797 & 2.241750 & $5.28 \pm 1.05$ & $1.51 \pm 1.53$ & $1.80 \pm 1.01$ & $2.80 \pm 0.69$ & $14.71 \pm 0.87$ & $10.35_{-0.13}^{+0.11}$ & $0.90_{-0.59}^{+0.79}$ & $40.61_{-21.24}^{+64.47}$ & $8.38_{-0.10}^{+0.09}$ \\
1893 & 235.720995 & 38.845859 & 2.244002 & $7.80 \pm 1.46$ & $1.06 \pm 2.50$ & ... & $3.23 \pm 1.07$ & $13.36 \pm 1.32$ & $10.33_{-0.08}^{+0.07}$ & $0.56_{-0.40}^{+0.65}$ & $35.05_{-14.12}^{+41.47}$ & $8.49_{-0.10}^{+0.09}$ \\
2702 & 235.705996 & 38.866214 & 2.255019 & $5.53 \pm 1.21$ & $3.35 \pm 2.31$ & $3.63 \pm 1.63$ & $1.60 \pm 1.14$ & $15.02 \pm 1.37$ & $10.29_{-0.05}^{+0.11}$ & $0.76_{-0.54}^{+0.82}$ & $33.74_{-16.47}^{+57.62}$ & $8.35_{-0.11}^{+0.11}$ \\
3531 & 235.693722 & 38.893516 & 2.269391 & $2.48 \pm 0.62$ & $0.02 \pm 0.81$ & $1.37 \pm 0.58$ & $1.65 \pm 0.48$ & $8.63 \pm 0.64$ & $10.25_{-0.13}^{+0.07}$ & $1.36_{-0.86}^{+0.97}$ & $39.85_{-26.15}^{+87.04}$ & $8.38_{-0.13}^{+0.11}$ \\
\noalign{\smallskip}\hline\noalign{\smallskip}
\multicolumn{13}{c}{Mass bin 1}   \\
\noalign{\smallskip}\hline\noalign{\smallskip}
2330 & 190.905581 & 35.926178 & 2.294822 & $6.85 \pm 0.78$ & $0.56 \pm 1.01$ & $0.97 \pm 0.74$ & $5.81 \pm 0.60$ & $25.95 \pm 0.80$ & $10.38_{-0.10}^{+0.07}$ & $1.94_{-0.60}^{+0.54}$ & $248.76_{-127.35}^{+221.17}$ & $8.43_{-0.08}^{+0.06}$ \\
3155 & 190.842877 & 35.957764 & 2.317641 & $6.76 \pm 1.31$ & ... & $1.38 \pm 1.41$ & $4.12 \pm 0.97$ & $12.65 \pm 1.20$ & $10.37_{-0.13}^{+0.13}$ & $0.84_{-0.54}^{+0.72}$ & $56.47_{-28.36}^{+76.80}$ & $8.55_{-0.10}^{+0.08}$ \\
3276 & 190.857451 & 35.964185 & 2.207442 & $17.43 \pm 1.79$ & $8.21 \pm 3.44$ & $1.58 \pm 1.88$ & $5.41 \pm 1.26$ & $32.71 \pm 1.45$ & $10.38_{-0.09}^{+0.07}$ & $0.25_{-0.18}^{+0.33}$ & $45.71_{-9.94}^{+23.24}$ & $8.40_{-0.06}^{+0.06}$ \\
5861 & 220.422593 & 40.054581 & 2.340491 & $5.14 \pm 1.60$ & $0.71 \pm 1.94$ & $0.96 \pm 1.38$ & $3.68 \pm 1.14$ & $15.35 \pm 2.10$ & $10.36_{-0.19}^{+0.18}$ & $1.42_{-0.90}^{+1.08}$ & $90.98_{-61.67}^{+224.83}$ & $8.42_{-0.16}^{+0.13}$ \\
1705 & 235.714718 & 38.841504 & 2.242345 & $3.74 \pm 0.99$ & ... & $1.24 \pm 0.89$ & $1.61 \pm 0.62$ & $7.68 \pm 0.77$ & $10.47_{-0.11}^{+0.13}$ & $0.83_{-0.58}^{+0.95}$ & $23.17_{-11.93}^{+45.72}$ & $8.45_{-0.12}^{+0.10}$ \\
3214 & 235.706084 & 38.884699 & 2.173583 & $3.50 \pm 0.61$ & $1.30 \pm 0.75$ & $0.37 \pm 0.52$ & $3.26 \pm 0.38$ & $23.76 \pm 0.51$ & $10.47_{-0.10}^{+0.07}$ & $1.84_{-0.95}^{+0.85}$ & $112.28_{-74.50}^{+184.52}$ & $8.20_{-0.11}^{+0.09}$ \\
3305 & 235.689707 & 38.887214 & 2.311889 & $3.61 \pm 1.09$ & ... & ... & $1.91 \pm 0.83$ & $9.23 \pm 1.04$ & $10.40_{-0.13}^{+0.12}$ & $1.18_{-0.79}^{+1.09}$ & $39.82_{-24.80}^{+100.93}$ & $8.41_{-0.15}^{+0.13}$ \\
5585 & 235.700570 & 38.959942 & 2.234094 & $10.48 \pm 1.48$ & $4.15 \pm 2.67$ & $2.41 \pm 1.62$ & $5.50 \pm 1.00$ & $18.64 \pm 1.21$ & $10.50_{-0.05}^{+0.06}$ & $0.57_{-0.38}^{+0.52}$ & $53.33_{-20.40}^{+46.83}$ & $8.53_{-0.08}^{+0.07}$ \\
5586 & 235.700082 & 38.960015 & 2.235445 & $6.64 \pm 1.38$ & $3.20 \pm 1.80$ & ... & $2.19 \pm 0.92$ & $5.96 \pm 1.14$ & $10.42_{-0.04}^{+0.05}$ & $0.37_{-0.28}^{+0.58}$ & $20.50_{-7.20}^{+18.97}$ & $8.61_{-0.10}^{+0.09}$ \\
5947 & 235.678266 & 39.007549 & 2.225628 & $3.89 \pm 1.05$ & $1.82 \pm 1.38$ & ... & $2.37 \pm 0.74$ & $12.52 \pm 0.89$ & $10.40_{-0.10}^{+0.08}$ & $1.28_{-0.83}^{+1.00}$ & $48.42_{-30.89}^{+108.11}$ & $8.36_{-0.13}^{+0.12}$ \\
\noalign{\smallskip}
\enddata
\tablecomments{All uncertainties reflect $1\sigma$ confidence intervals. Missing or undetected emission lines are indicated by ellipses (...), 
representing lines that are either below the detection threshold or masked due to contamination. }

\label{tab:indvd}
\end{deluxetable}
}
\newpage
\section{HST Observed Layouts}
\begin{figure}[!h]
    \centering
    \includegraphics[width=1\linewidth]{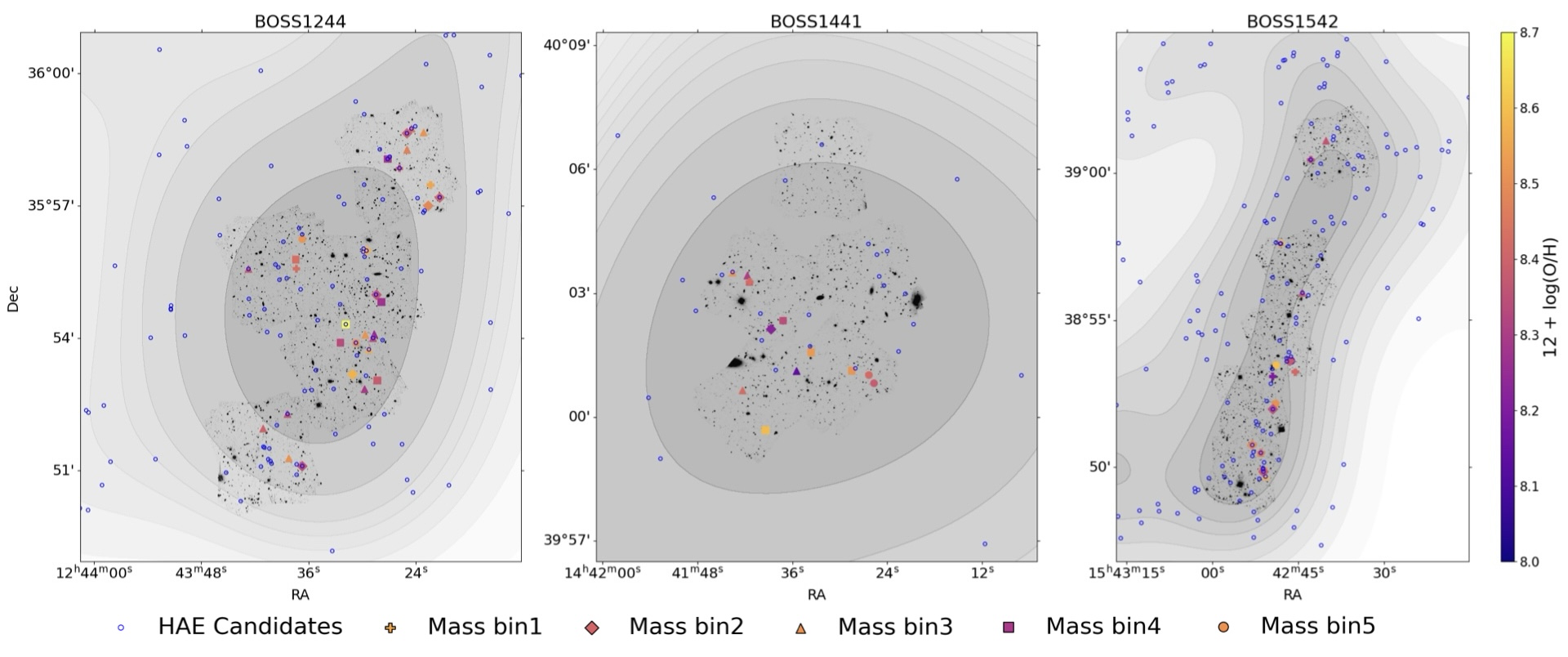}
    \caption{\yy{Layouts and observed fields of the \textit{HST}/WFC3 F125W imaging for the three protoclusters. Individual sample galaxies are shown with different symbols representing their stellar-mass bins and colors indicating their inferred metallicities. The background contours show the projected density maps of HAEs (for BOSS1244 and BOSS1542) and LAEs (for BOSS1441), with darker shades corresponding to higher densities. All observed fields are centered on the core regions of their respective protoclusters.}}
    \label{fig:protoclusters}
\end{figure}
\newpage
\section{SED fitting examples}
\begin{figure}[!h]
    \centering
    \includegraphics[width=1\linewidth]{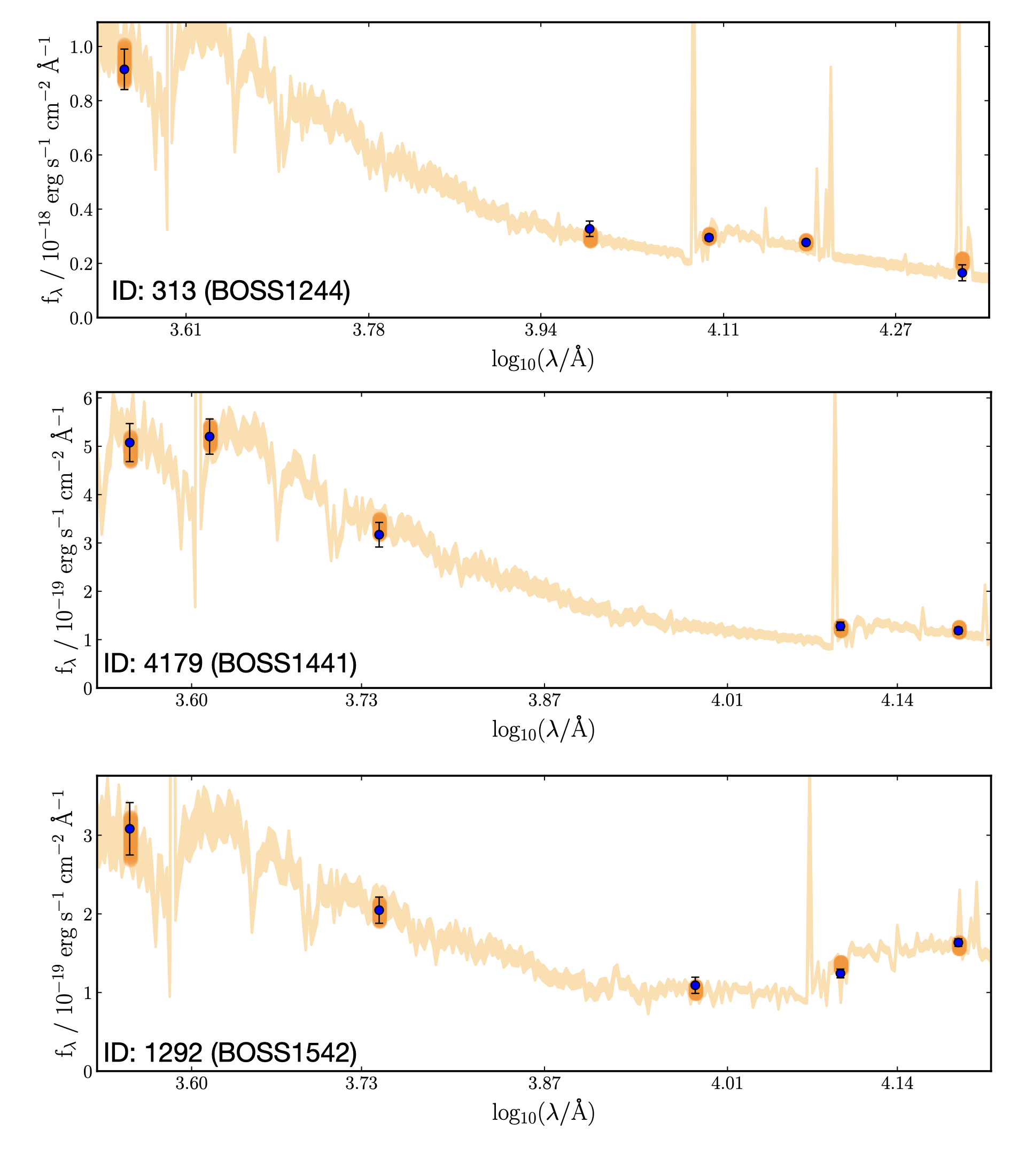}
    \caption{\yy{Examples of SED fitting for individual galaxies in the three protoclusters.
     Each panel shows the observed photometry (data points with error bars) and the best-fit model SED (solid line) along with the 1$\sigma$ uncertainty range (shaded region).}}
    \label{fig:SED_fitting_cases}
\end{figure}

\newpage

\bibliography{sample631}{}
\bibliographystyle{apj}



\end{CJK*}
\end{document}